\documentclass[11pt, a4]{article}
\usepackage{amssymb,amsmath,latexsym,graphicx, bm, mathrsfs}
\usepackage[hidelinks,citecolor=blue,urlcolor=blue,colorlinks=true,bookmarks=false,hypertexnames=true]{hyperref} 
\usepackage{natbib} %Bibliography
\setcitestyle{numbers,square} %Cite as numbers or author-year.
\usepackage{graphics,graphicx, cancel}
\usepackage[dvipsnames]{xcolor}
\usepackage{graphicx}
\graphicspath{ {./images/} }
\usepackage[british]{datetime2}
\usepackage[page,toc,titletoc,title]{appendix}
\usepackage{tocloft}
\usepackage{blindtext}
\usepackage{dsfont}
\usepackage{dirtytalk}

\DeclareMathOperator{\cosec}{cosec}
\newcommand\numberthis{\addtocounter{equation}{1}\tag{\theequation}}
\usepackage{comment}
\usepackage{caption}
\usepackage{enumitem}
\usepackage{mathrsfs}

\begin{document}

\title{Harvesting Fermionic Field Entanglement in Schwarzschild Spacetime
\vspace{-0.25cm} 
\author{Nitesh K. Dubey$ ^{a,b}$, Sanved Kolekar$^{a,b}$ \\ \vspace{-0.5cm} \\ 
\textit{$ ^a $Indian Institute of Astrophysics} \\ 
\textit{Block 2, 100 Feet Road, Koramangala,} 
\textit{Bengaluru 560034, India.} \\ 
\textit{$ ^b $Pondicherry University} \\ 
\textit{R.V.Nagar, Kalapet, Puducherry-605014, India}\\
\texttt{\small Email: \href{mailto:nitesh.dubey@iiap.res.in}{nitesh.dubey@iiap.res.in}, 
\href{mailto:sanved.kolekar@iiap.res.in}{sanved.kolekar@iiap.res.in}}}}

\maketitle

\abstract{ We explore entanglement harvesting using two Unruh-DeWitt (UDW) detectors linearly coupled to the scalar density of a massless spin-1/2 field in 1+1 Schwarzschild spacetime. We consider different vacua, including the Boulware, Hartle-Hawking-Israel (HHI), and Unruh vacua, and investigate various configurations of detector trajectories. We find that the transition rate of the static UDW detector exhibits the expected Planckian behavior in the HHI state, while the Unruh state leads to the Helmholtz free energy density of a fermionic thermal bath. We demonstrate that the near-horizon entanglement properties for static detectors in the HHI state have similar behaviour to those in Minkowski vacua for uniformly accelerated detectors in Rindler spacetime. We further consider a different interaction  Hamiltonian which breaks local Lorentz symmetry and find that the transition rate of the static detector still exhibits Planckian behavior in the HHI state, while in the Unruh state, it leads to the Helmholtz free energy density of a bosonic or fermionic thermal bath corresponding to the static or conformal 2-bein in interaction, respectively. We observe that the anti-Hawking effect enhances the entanglement between the two detectors while the gravitational redshift and Hawking radiation decrease it. In particular, due to the presence of the anti-Hawking effect, the mutual information and concurrence near the event horizon can be non-zero even for static detectors with static 2-bein, which is in contrast with the case of the scalar field. Conclusions are discussed. 
}

\pagebreak

\tableofcontents

\pagebreak

\section{Introduction}
The black hole information loss problem has been explored from various perspectives, such as probabilistic spacetime theory,  transfinite set theory, quantum metrology, quantum teleportation, relativistic quantum information, the AdS/CFT correspondence,  and more \cite{pst, tst, Lowe:1999pk, Raju:2020smc}. One of the crucial elements in addressing the black hole information loss problem is the quantum entanglement present in matter fields. To have a deeper understanding of the entanglement inherent in the field, one can consider the associated algebra and its properties in the context of a freely chosen tensor-product decomposition of the full state Hilbert space \cite{Zanardi}. The Hilbert space decomposition depends upon the reference frames that one chooses. Furthermore, the curvature of spacetime also plays a significant role by affecting the derivatives in field equations that influence the degree of entanglement within the field \cite{martinl, Saravani, Kempf}. The entanglement dynamics of spacetime and matter are particularly important in the framework of a information complete quantum field theory, which may pave a way for unifying quantum theory and general relativity \cite{Faulkner:2022mlp, Chen:2014zoo}. Additionally, entanglement and other vacuum correlations has also been shown to be affected by other elements, such as the Hawking and local Unruh effect \cite{tjoaman, mgyhm, gytm, PhysRevD.111.065004}.  These effects are also used to explain the transformation between different forms of vacuum correlations \cite{Zhang:2009jn, Giddingsqit}. The expected emergence of Hawking particles (or quasi-particles) also carries substantial significance at various other places \---\ apart from black hole physics\---\ in the natural phenomena where quantum excitations propagate within a Lorentzian geometry resembling black holes \cite{carlos}. 

In the pursuit of harvesting the entanglement, \cite{valentini} initially and later \cite{reznik, PhysRevLett.88.230402, Peres:2002wx, PhysRevLett.91.180404}  proposed a way to harvest the entanglement using quantum probes. 
In particular, \cite{reznik} investigated the entanglement present in the vacuum of a massless scalar field by using a pair of causally disconnected two-level detectors coupled to the field. Subsequently, different types of probes were analyzed, including quantum harmonic oscillators and qubits, employing various coupling types, including both linear and nonlinear couplings, in various spacetimes \cite{PhysRevD.105.125011, rick, rick2, PhysRevD.109.045018, deSouzaLeaoTorres:2024vrl, Ruep:2021fjh, PhysRevD.105.065016, Barman:2021kwg, Chowdhury:2021ieg}.  In particular, \cite{rick} and \cite{rick2} introduce the formalism and limitations of a covariant model of smeared particle detectors. \cite{PhysRevD.105.065016} use a complex detector model to study the antiparticle sector of a non-Hermitian field.  The distinct entanglement-harvesting profiles due to minimal and conformal couplings have been illustrated in \cite{nambu}. \cite{Alejandro} showed that vacuum entanglement harvesting is significantly more efficient for a smooth switching mechanism than a sudden one, especially for spacelike separations. Moreover, the performance of measurements on quantum fields also affect entanglement harvesting \cite{PhysRevD.107.045011}.

In the context of black holes, the entanglement harvesting protocol described in \cite{tjoaman, gytm, Henderson:2017yuv}, involving static detectors coupled to a real massless scalar field, suggests that the horizon inhibits any kind of correlation near it. However, freely falling detectors coupled to the massless scalar field don't experience such decay due to the finite relative gravitational redshift \cite{gytm}.  Furthermore, at near infinite acceleration of the quantum probes, the entanglement of the fermionic field can be either zero or nonzero, unlike that of the scalar field, depending on the choice of the state in the Rindler spacetime as well in the near horizon limit of a Schwarzschild black hole \cite{Montero:2011sx, Martin-Martinez:2009hfq}. The generation of entanglement between modes pair of a quantum field within a single, rigid cavity undergoing non-uniform motion in Minkowski space-time is also shown to be influenced by the field's statistics \cite{Friis:2012tb}.  Moreover, the entanglement tradeoff across horizons also depends on the statistics \cite{PhysRevA.81.032320}. The entanglement properties of fermionic fields differ from those of bosonic fields in several contexts. Notably, the manner in which fermionic fields become entangled prompted \cite{PhysRevD.82.045030} to suggest that fermionic field entanglement, arising from the expansion of the universe, can be used to reconstruct parameters characterizing the history of cosmic expansion.

Despite the certain different entanglement properties of fermionic and bosonic fields, studies on the entanglement harvesting protocol involving a detector moving along a specified trajectory and coupled to a fermionic field in a black hole spacetime are limited. In particular, the roles of gravitational redshift, the Hawking and anti-Hawking effects, vacuum polarization, and similar phenomena have not been extensively explored for fermionic fields. Takagi \cite{takagi} initially explored the detector response function of a Unruh DeWitt (UDW) detector coupled linearly to the scalar density of a massless Fermionic field in Minkowski spacetime. Subsequently, \cite{jorma} and \cite{mokhtar} demonstrated the similarities and differences between the UDW coupled linearly to the scalar density of the massless fermionic field and that coupled to a massless scalar field in flat spacetime. Notably, while the numerical values of the entanglement measures, harvested using detectors, depend on various detector parameters --- such as the energy gap and the switching function --- a systematic comparison can still be made by keeping certain parameters fixed while varying others. \cite{PhysRevD.105.065016} studied an example of entanglement harvesting from a fermionic field in flat spacetime. However, the study of entanglement harvesting for fermionic fields in curved spacetime is limited. This further motivates us to compare an analysis of entanglement harvesting for a detector coupled to the scalar density of the Fermionic field with that of a scalar field in the background of a Schwarzschild black hole.  

In section \ref{section2}, we start with a concise review of quantum field theory applied to fermionic fields within the context of curved spacetime and then present the derivation of the two-point functions needed in subsequent sections in various vacuum states. The two-point function in a pure global Gaussian state contains all information about the correlations of the field. We further briefly review the calculation of the relative entanglement entropy of a massless fermionic field for disjoint intervals in the Schwarzschild spacetime using the two-point function. This yields various interesting properties and inequalities associated with the mutual information of the field in various states. Section \ref{section3} introduces the Unruh DeWitt detector (UDW) formalism for fermionic fields and calculates the transition rates of the quantum probe in different vacua, as the transition probability is a competing term in the correlation measures. The transition probability in the usual scenario is positively correlated with temperature; however, it can also decrease with an increase in temperature due to the anti-Hawking effect \cite{antihawking}. We introduce the condition for the anti-Hawking effect and later correlate it with results obtained from entanglement harvesting. Section \ref{section4} starts with a discussion of the entanglement measures used in the later subsections. We investigate entanglement harvesting with two static detectors—one in close proximity to the horizon and the other at some proper distance from it. We vary the proper distance between detectors, keeping other parameters constant, and perform entanglement harvesting. The subsequent subsection explores the entanglement characteristics of the vacuum relative to the distance from the horizon, with fixed detector separation and other fixed parameters. The same analysis is also repeated for both detectors in free-fall trajectories in the following subsection. Finally, section \ref{section5} provides a comparative analysis of the near-horizon entanglement properties of the Hartle-Hawking state with those of the Minkowski vacua in the context of a uniformly accelerated detector in flat spacetime. Conclusions are discussed in the last section. Use is made of natural units, namely $\hbar$=c= $k_B $=1, throughout the paper.

\section{QFT in spherically symmetric spacetime } \label{section2}

We begin with a concise review of quantum field theory applied to fermionic fields within the context of curved spacetime and then present the derivation of the two-point functions in various vacuum states. 

We define spacetime as a pair ($\mathcal{M}$, g), where $\mathcal{M}$ is a connected 4-dimensional Hausdorff manifold and g is a metric of signature +2 on  $\mathcal{M}$ \cite{hawking}. If it admits an isometric action $\Phi: SO(3) \times  \mathcal{M} \rightarrow \mathcal{M}  $ such that the maximal dimensions of its orbits are two, then we call the spacetime to be spherically symmetric. Any spherically symmetric solution of the vacuum Einstein field equation exhibits local isometry with the Schwarzschild solution \cite{bergmann}. In particular, a non-rotating body undergoing gravitational collapse results in the formation of a Schwarzschild black hole. In this paper, we focus specifically on the Schwarzschild spacetime.

\subsection{The classical Schwarzschild spacetime}
The Schwarzschild spacetime possesses three globally spacelike Killing vector fields, in addition to one more Killing vector field that is timelike outside the horizon and spacelike inside the horizon. The metric in (1+3) dimensions, using Schwarzschild coordinates, can be expressed as follows:
\begin{equation} \label{eq:1}
    ds^2 = -\bigg(1 - \frac{2GM}{r} \bigg) dt_S ^2 + \frac{1}{\bigg(1 - \frac{2GM}{r} \bigg)} d r^2  + r^2 (d \theta ^2 + \sin ^2{\theta} d \phi^2 ) .
\end{equation}
Here $M$ is the ADM mass associated with the geometry, and $ \{ t_S,r,\theta ,\phi \}$ represents the Schwarzschild coordinates. Furthermore, one should also note that the above expression, denoted by Eq.(\ref{eq:1}), is a classical relation and does not say anything about the quantum states of space-time. However, one can take various matter fields and study their quantum theory in a classical curved spacetime.

\begin{figure*}[ht!] 
        \centering
        \includegraphics[width=0.72\textwidth]{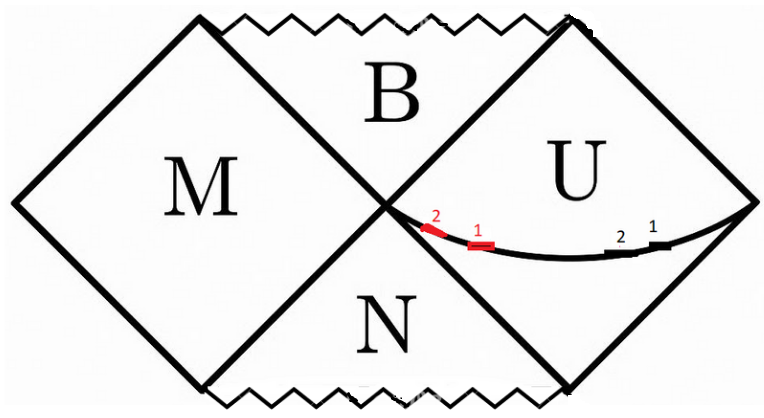}
        \captionsetup{margin=1cm, font=small}
        \caption{The above plot shows the Penrose diagram for a Schwarzschild black hole. We study the correlation measures of the massless Dirac field in region U. We show four disjoint intervals on a constant Schwarzschild time slice to illustrate the computation of the correlation measures between the red 1 and black 1 intervals, and then between the red 2 and black 2 intervals, corresponding to various distances from the event horizon, while keeping the proper distance between intervals fixed.}    
        \label{fig:0}
\end{figure*}

For ease of application in later sections, let us introduce null coordinates v = $t_S + r_*$, u = $t_S - r_*$, with $r_* = r + R_s \log |r/R_s -1|$ and $R_s= 2GM$. Essentially, the constant value of any one of these coordinates describes null geodesics corresponding to ingoing and outgoing directions, respectively. While certain entanglement properties depend upon the angular components, the majority of these properties remain valid when we confine our focus to radial motion exclusively.  The truncated Schwarzschild line element, in (1+1) dimensions, in the new coordinates becomes
\begin{equation} \label{eq:2}
   ds^2 = - \frac{R_s}{r} e^{-r/R_s} e^{(v-u)/2R_s} du dv .
\end{equation}
The above metric, represented as Eq.(\ref{eq:2}), has a coordinate singularity at $R_s$ = $2GM$. Nonetheless, it is possible to perform an analytic continuation of this metric, making it regular at all points except for r=0. This can be achieved by introducing new coordinates, $ U := \mp $ $ 2 R_s \exp{(-u/2 R_s)}$ and $ V :=  $ $2 R_s \exp{(v/2 R_s)}$, which are commonly referred to as the Kruskal coordinates. The line element corresponding to Kruskal coordinates is
\begin{equation} \label{eq:3}
   ds^2 = - \frac{R_s}{r} e^{-r/R_s } dU dV.
\end{equation}
In terms of U and v one obtains,
\begin{equation} \label{eq:4}
   ds^2 = - \frac{R_s}{r} e^{-\frac{r}{R_s} + \frac{v}{2 R_s}} dU dv .
\end{equation}
\subsection{Dirac field in Schwarzschild spacetime}
There may exist a diverse variety of fields on $\mathcal{M}$ that describe the matter content of the spacetime, adhering to either tensorial or spinor equations, and are defined using the metric g on $\mathcal{M}$. For example, we can have electromagnetic, neutrino, electron, Higgs, or pion field. Metrics other than g or other connections can also be understood as other physical fields. Spinors and spinorial tensor fields arise as realizations of the representation of ISL(2,$\mathbf{C}$). However, one can define the spinor fields only if the spacetime manifold satisfies certain topological properties \cite{wald}. The matter field we consider in the Schwarzschild spacetime is the Dirac spinor, which comprises four components. One can understand it as the combination of an SL(2,$\mathbf{C}$) spinor and its complex conjugate SL(2,$\mathbf{C}$) spinor counterpart. The action describing the Dirac field is given by 
\begin{equation} \label{eq:5}
  \mathcal{A} =  \int d^4x  \sqrt{-g(x)} \{\frac{1}{2}i [\Bar{\psi} \tilde{\gamma} ^ \mu \nabla _\mu \psi - (\nabla _\mu \bar{\psi} )\tilde{\gamma} ^ \mu \psi  ] - m \bar{\psi} \psi \} ,
\end{equation}
where $\nabla$ represents the spinor covariant derivative and $\tilde{\gamma} ^ \mu$ are the curved space Dirac gamma matrices. One can get the following Dirac equations by varying the above action, denoted by Eq.(\ref{eq:5}) with respect to $\bar{\psi}$ and $\psi$, respectively:
\begin{equation*} \numberthis \label{eq:6}
     (i \tilde{\gamma}^\mu \nabla _\mu - m ) \psi = 0,
\end{equation*}
\begin{equation*}
    i \nabla _\mu \bar{\psi} \tilde{\gamma}^\mu  + m  \bar{\psi} = 0 .
\end{equation*}
In n-dimensional spacetime, the above Eqs.(\ref{eq:5}-\ref{eq:6}) exhibit conformal invariance in the massless limit as long as  $\psi \rightarrow \Omega ^{(1-n)/2} (x) \psi$ under conformal transformation of the metric, $g_{\mu \nu} (x) \rightarrow  \Omega ^2(x)  g_{\mu \nu} (x)$. Under the combined local Lorentz and general coordinate transformations on Eq.(\ref{eq:5}), by imposing the requirement of general covariance of the Dirac equations, Eq.(\ref{eq:6}), one obtains a condition that is identical to the requirement of covariance under Lorentz transformations in flat spacetime. The key distinction is that in this context, the transformations are local \cite{parker} while analyzing quantum entanglement one has to look for  nonlocal correlations. Solutions of the Dirac equation form a complete and orthonormal set of modes that can be quantized by imposing anticommutation relations on the field because quantizing the Dirac field using commutators leads to issues with causality and the energy not being bounded from below \cite{parker}. 

The Schwarzschild metric depends upon the radial coordinate, `$r$', which becomes timelike inside the horizon at $r = R_s$ = $2 GM$. Furthermore, the Schwarzschild spacetime exhibits time-reversal symmetry; however, the quantum state imposed on it need not share the same time-reversal symmetry. However, even with the time dependence of the interior region of the Schwarzschild solution, it is possible to define a globally stable vacuum state \cite{boulware}.  Since the manifold is time-oriented, one can also define the Feynman propagator, which sandwiches time-ordered quantum fields between the state, as follows:
\begin{eqnarray} \label{eq:7}
    i S^F _{\alpha \beta} (x,x') = \langle 0|T (\psi_\alpha (x) \bar{\psi}_\beta (x'))|0 \rangle
                = \theta (t-t') S_{\alpha \beta} ^ + (x,x') - \theta (t'-t) S_{\alpha \beta} ^ - (x,x'),
\end{eqnarray}
where $S_{\alpha \beta} ^ +$ and $S_{\alpha \beta} ^ -$ are known as Wightman bi -distributions. The Wightman bi-distributions have a Dirac delta form for null-separated events \footnote{The Wightman bi-distribution gives well-defined results when one integrates over appropriate regions with well-defined boundary conditions. Therefore, we use the term Wightman function in the rest of the paper.}. However, due to relativistic causality, since the Dirac field anticommutes for spacelike-separated events, the time-ordered product does not exhibit any discontinuity when $x_0 = x'_0$. The Feynman propagator for the spin half field satisfies
\begin{equation} \label{eq:8}
    [i \tilde{\gamma} ^\mu (x) \nabla _\mu ^x - m ] S^F (x,x') = [-g(x)]^{-1/2} \delta ^n (x-x').
\end{equation}
It can be seen by direct substitution that
\begin{equation} \label{eq:9}
     S^F (x,x') =  [i \tilde{\gamma} ^\mu (x) \nabla _\mu ^x + m ] G_\phi (x,x'),
\end{equation}
where $G_\phi (x,x')$ is a bispinor with each component satisfying the Green's function equation for Klein-Gordon field in curved space-time with a nonminimal coupling \cite{birrell}. It can be verified using the identity shown in Eq.(3.232) of \cite{parker}. In the subsequent sections, we will work with a spacetime that has a zero Ricci scalar, so the Ricci scalar term will be absent. Writing $G_\phi (x,x')$ in terms of scalar Wightman function, $ W_\phi (x,x'), $ we have \cite{Muck2000, Allen1986, Letsios:2020twa, Avramidi2000}
\begin{equation} \label{eq:10}
    i S^F _{\alpha \beta} (x,x') =  \theta (t-t') [(i \tilde{\gamma} ^\mu (x)  \nabla _\mu ^x + m )U(x,x')  W_\phi (x,x') ]_{\alpha \beta}  + \theta (t'-t) [(i \tilde{\gamma} ^\mu (x)  \nabla _\mu ^x + m )U(x,x') W_\phi (x',x)]_{\alpha \beta} .
\end{equation}
Here, \( U(x,x') \) represents the spinor parallel propagator, which ensures that the above equation, expressed in terms of the scalar field Wightman function, transforms identically to Eq.~\eqref{eq:7}\cite{Allen1986, Letsios:2020twa, Avramidi2000}. Again, relativistic causality ensures that the Feynman propagator \eqref{eq:10} is analytic at equal times, and furthermore, the first time derivative is also continuous at equal times. Comparing the above equation, Eq.(\ref{eq:10}), with Eq.(\ref{eq:7}), we get the following expressions for the Wightman function of the Dirac field:
\begin{eqnarray} \label{eq:11}
    S_{\alpha \beta} ^ + (x,x') = [(i \tilde{\gamma} ^\mu (x)  \nabla _\mu ^x + m )U(x,x') W_\phi (x,x')]_{\alpha \beta} ,
&&  S_{\alpha \beta} ^ - (x,x') = - [(i \tilde{\gamma} ^\mu (x)  \nabla _\mu ^x + m) U(x,x') W_\phi (x',x) ]_{\alpha \beta}.
\end{eqnarray}
In (1+1)-dimensional Minkowski spacetime, the Wightman function for the Dirac field at two points $\tilde{x}$ and $\tilde{x}'$ on a constant time slice is given by
\begin{eqnarray} \label{eqn:12}
    S^+ (\tilde{x},\tilde{x}') = \int \frac{dp}{2\pi} \frac{p_\mu \gamma^\mu + m}{2 \sqrt{p^2 + m^2}} \gamma^0 e^{- i p (\tilde{x} - \tilde{x}')},
\end{eqnarray}
which evaluates to the following expression in the massless limit:
\cite{parker, Casini:2009sr, Casini:2009vk, Guo:2022ivd}
\begin{eqnarray} \label{eqn:13}
    S ^ + (\tilde{x},\tilde{x}') = \frac{1}{2} \delta (\tilde{x}-\tilde{x}') \mathds{1} + \frac{i}{2 \pi} \mathcal{P} \bigg( \frac{1}{\tilde{x}-\tilde{x}'} \bigg) \gamma^0 \gamma^1 .
\end{eqnarray}
Here, $\mathds{1}$ is the identity matrix, and $\mathcal{P}$ represents that the principal value regularization has been applied. Using the property of conformal invariance, discussed above, we get the following Wightman function for the massless Dirac field in any spacetime which is conformally related to the (1+1) D Minkowski spacetime by a conformal factor $\Omega$ \cite{Guo:2022ivd}:
\begin{eqnarray} \label{eqn:14}
    S_{\Omega^2 g} ^ + (\tilde{x},\tilde{x}') = \Omega^{-1/2} (\tilde{x}) \Omega^{-1/2} (\tilde{x}') \bigg(\frac{1}{2} \delta (\tilde{x}-\tilde{x}') \mathds{1} + \frac{i}{2 \pi} \mathcal{P} \bigg(  \frac{1}{\tilde{x}-\tilde{x}'} \bigg) \gamma^0 \gamma^1  \bigg).
\end{eqnarray}
One can also get the above Wightman function, Eq.\eqref{eqn:14}, from Eq.\eqref{eq:11} by using $\tilde{\gamma}^\mu = \Omega(x)^{-1} \gamma ^\mu$, $U(x,x')= \sqrt{\Omega(x)/\Omega(x')} U_M(x,x')$ where $U_M(x,x')$ is the $1+1$ Minkowski spinor parallel propagator and $\gamma ^\mu \nabla_\mu U(x,x') = 0$.
\subsection{Different vacuua}

One usually defines a vacuum state as a Gaussian pure state. The term `pure' means that one cannot decompose the state into a convex combination of other states, and the Gaussian states are states that can be fully described by their one and two-point correlation functions. Generally, it's not anticipated that the vacuum should be devoid of particles for observers following distinct worldlines. Nevertheless, the symmetries inherent in spacetime dictate the emergence of certain mode sets as natural choices for specific observers. Guided by the principles of normalization and equivalence, one opts for Hadamard states, which are physical states that exhibit a resemblance to the Minkowski vacuum in the high-energy ultraviolet (UV) limit. In a stationary spacetime with a bifurcate Killing horizon, one can identify at most one stationary Hadamard vacuum (see theorem 5.1 of \cite{kay}). One can refer \cite{paul} for the detailed discussion of states using massless Dirac fields in the Schwarzschild spacetime. For ease of application in the later sections where we use the detectors coupled to the scalar density of the massless fermionic field, this section delves into an exploration of three distinct vacuum states associated with the Schwarzschild spacetime \footnote{We note that only one is Hadamard over the maximally extended Schwarzschild spacetime. Moreover, uniqueness does not necessarily imply existence in a more general scenario \cite{kay}.} and their corresponding two-point functions
\begin{align}
    W_{\psi} ^\alpha (x,x')  & = \langle 0_\alpha|:\bar{\psi}_a (x) \psi _a (x) : :\bar{\psi}_b (x') \psi _b (x') : |0_\alpha \rangle \label{deftwopoint} \\ 
    & = - N  g ^ {\mu \rho} (x)  \Lambda ^\nu _\rho (x',x) \partial _\mu ^x W_\phi ^\alpha  (x,x')  \partial _\nu ^{x'} W_\phi ^\alpha  (x,x') \label{generaltwoo}
\end{align}
Here, :: denotes normal ordering, $\alpha$ labels different states, $N$ is the number of spacetime dimensions, $\Lambda ^\nu _\rho $ represents the vector parallel transport, and $W_\phi^\alpha(x, x')$ is the Wightman function of a massless scalar field in the corresponding vacuum. The above expression \eqref{generaltwoo} is invariant under general coordinate transformation as well as local Lorentz transformation, and it has been derived in Appendix [\ref{Appendix A}].

The local Lorentz invariance is crucial in defining horizons and the notion of a black hole \cite{Barausse:2013nwa}. It has been suggested in \cite{Jacobson:2010fat} that Lorentz symmetry violations could allow perpetual motion machines of the second kind, and it is proposed that Lorentz symmetry may emerge macroscopically from a microscopic second law of causal horizon thermodynamics. Further, \cite{PhysRevD.84.124043} suggests that the violation of Lorentz invariance could allow processes violating the second law of thermodynamics, which is related to the unitarity of the underlying microscopic theory. One can refer \cite{PhysRevD.88.096006} for a brief study of fermions in the presence of Lorentz violations. In our case of interaction Hamiltonian giving the two-point function shown in Eq.\eqref{generaltwoo}, the Lorentz invariance is preserved due to the presence of the vector parallel propagator $\Lambda^\nu_\mu$. In order to see the effect of violating local Lorentz invariance, with a toy model interaction, we consider a Lorentz symmetry-violating interaction. By defining the decomposition of the spinor parallel propagator as \( U(x',x) = \bar{A}(x') A(x), \quad U(x,x') =  \bar{A}(x) A(x') \) at every spacetime point, and rescaling the spinor fields as  \(
\psi'(x) = A(x) \psi(x), \quad \bar{\psi}'(x) = \bar{A}(x) \bar{\psi}(x), \) we get an interaction Hamiltonian, along with the following corresponding two-point function computed in Appendix~[\ref{Appendix A}]:
\begin{align} 
 W_{\psi,b} ^{'\alpha} (x,x') & = \langle 0_\alpha|  :\bar{\psi}'_a (x) \psi' _a (x) :  :\bar{\psi}'_b (x') \psi' _b (x') : |0_\alpha \rangle \label{deftwopointold} \\
 & = - N b_\delta ^{.\mu} (x) b_\beta ^{.\nu} (x') \eta ^ {\delta \beta} \partial _\mu ^x W_\phi ^\alpha  (x,x')  \partial _\nu ^{x'} W_\phi ^\alpha  (x,x') . \numberthis \label{generaltwooold}
\end{align}
The index $b$ represents the choice of the two being taken as the two-point function \eqref{generaltwooold} now depends on the choice of 2-bein $b_\delta ^{.\mu} (x)$.

In the rest of the paper, we follow the notation that $W_{\psi} ^\alpha (x,x')$ without the prime is derived from the interaction Hamiltonian $\hat{H} _{\text{j}}^{\text{int}}$ in Eq.(\ref{eq:24}) and is independent of the choice of 2-bein while $W_{\psi,b} ^{'\alpha} (x,x')$ with the prime is derived from the interaction Hamiltonian $\hat{H} _{\text{j}}^{'\text{int}}$ in Eq.(\ref{eq:24old}) and is dependent of the choice of 2-bein.

\subsubsection{HHI vacuum}
In the eternal Schwarzschild spacetime, there exists exactly one stationary Hadamard vacuum state, which is defined over the entire maximally extended spacetime, known as the Hartle-Hawking-Israel (HHI) state \cite{israel}. One understands the left and right exteriors of the event horizon, in this state, to be connected by an  Einstein-Rosen bridge, which makes them entangled. To define this state, one quantizes the field using modes in terms of positive frequency with respect to both past and future horizon generators $\partial _U $ and $\partial _V$. The energy-momentum tensor here has a relatively low magnitude \cite{visser3}, and the geometry closely resembles the classical Schwarzschild solution. Additionally, a static observer far from the event horizon experiences both outgoing and ingoing fluxes of radiation in thermal equilibrium.

The 1+1-dimensional Schwarzschild metric in terms of U and V coordinates (see Eq.~\eqref{eq:3}) is related to the Minkowski spacetime metric by a conformal factor as \( g_{\mu \nu}(x) = \Omega'^2(x) \eta_{\mu \nu}(x)\). Therefore, one can always choose a local orthogonal frame $e^a _\mu(x) = \Omega'(x) \delta ^a _\mu $ and compute the spin connection to yield the following vector parallel propagator:
\begin{equation} \label{hhiparall}
    \Lambda ^\mu _\nu (x,x') = \frac{\Omega'(x')}{\Omega'(x)} \delta ^\mu _\nu.
\end{equation}
Substituting the above expression \eqref{hhiparall} of the vector parallel propagator in the two-point function of the detector coupled linearly to the scalar density of a massless Dirac field, shown in Eq.\eqref{generaltwoo},  with the metric Eq.\eqref{eq:3}, we get the following two-point function:
\begin{equation} \label{confwighh}
    W_{\psi} ^{\text{HHI}} (x,x') = -\frac{\sqrt{rr'}}{2 \pi ^2 R_S} e^{(r+r')/2R_S} \frac{1}{(\Delta U - i \epsilon)(\Delta V - i \epsilon)}.
\end{equation}
Now, using the invariance of Eq.~\eqref {generaltwoo} under the local Lorentz transformations and the general coordinate transformations, one can say that Eq.~\eqref {confwighh} is valid in all reference frames. Since all two-dimensional metrics are related by conformal transformation, imposing the condition of conformal invariance of the action \eqref{eq:5} for a massless Dirac field (\(m = 0\)), for which the field should transform as \(\psi \rightarrow \Omega^{(1-n)/2}(x) \psi\) under a conformal transformation of the metric, \(g_{\mu \nu}(x) \rightarrow \Omega^2(x) g_{\mu \nu}(x)\), also allows one to get the same two-point function, Eq.\eqref{confwighh}, using the known two-point function in Minkowski spacetime. One can also get the same expression \eqref{confwighh} for the two-point function shown in Eq.\eqref{generaltwooold} by substituting 2-beins \footnote{We refer to the (1+1)-dimensional counterpart of a tetrad as the 2-bein.}
\begin{eqnarray} \label{conftetrh}
b_0^{\mu} = \left( \begin{array}{c}
  \frac{e^{(v-u)/4 R_s}}{\sqrt{1-R_s/r}} \\
  \frac{e^{(v-u)/4 R_s}}{\sqrt{1-R_s/r}}
\end{array} \right),
\quad
b_1^{\mu} = \left( \begin{array}{c}
  \mp \frac{e^{(v-u)/4 R_s}}{\sqrt{1-R_s/r}} \\
  \pm \frac{e^{(v-u)/4 R_s}}{\sqrt{1-R_s/r}}
\end{array} \right),
\end{eqnarray}
in Eq.\eqref{generaltwooold}. 

Since the two-point function shown in Eq.~\eqref{generaltwooold} depends on the choice of the 2-bein, we introduce a subscript \( b \), i.e., \( W_{\psi, b}^{'\text{HHI}} \), to indicate which 2-bein is being used, both here and throughout. So, in this notation, \( W_{\psi, c}^{'\text{HHI}} = W_{\psi}^{\text{HHI}} \), where \( c \) denotes the conformal tetrad. The timelike component of the 2-bein shown in Eq.\eqref{conftetrh} represents the four-velocity of a reference frame that is neither freely falling nor static (see Fig \ref{fig:0}). Converting the four-velocity in t-r plane by usual tetrad transformation and integrating one gets the following trajectory:
\begin{equation}
    \frac{r_*}{2 R_s} = - \ln{(\cosh{\frac{t}{2 R_s}})} + C_1,
\end{equation}
with proper acceleration
\begin{equation}
    a := \sqrt{a^\mu a_\mu} = \pm \frac{1 - R_s ^2 / r^2}{4 R_s \sqrt{1 - R_s/r}},
\end{equation}
where $r_* = r + R_s \ln{|r/R_s - 1|}$ is the usual tortoise coordinate and $C_1$ is an integration constant. For illustration purposes, we show the plots for trajectory and acceleration for a particular $C_1$ in Fig. \ref{fig:0}.

However, the choice of reference frame at any specific point in a curved manifold is entirely arbitrary. One can apply a local, spacetime-dependent Lorentz transformation to Eq.\eqref{conftetrh} to obtain a 2-bein describing the same geometry but whose zeroth component represents the four-velocity of a desired reference frame. The 2-bein corresponding to a static observer in the metric Eq.(\ref{eq:3}) is given by: 
\begin{eqnarray} \label{eq:12}
b_0^{\mu} = \begin{pmatrix}
  \frac{-U}{2R_s\sqrt{1-R_s/r}} \\
  \frac{V}{2 R_s\sqrt{1-R_s/r}}
\end{pmatrix},
\quad
b_1^{\mu} = \begin{pmatrix}
  \pm \frac{U}{2 R_s\sqrt{1-R_s/r}} \\
  \pm \frac{V}{2 R_s\sqrt{1-R_s/r}}
\end{pmatrix} .
\end{eqnarray}
The above 2-bein is related to Eq.\eqref{conftetrh} by just a local Lorentz transformation. Using the 2-bein, Eq.(\ref{eq:12}), in the  Eq.\eqref{generaltwooold}, we obtain the following two-point function for the scalar density of the Fermionic field in the HHI vacuum:
\begin{equation} \label{eq:13}
    W_{\psi,s} ^{'\text{HHI}} (x,x') = \frac{1}{16 \pi ^2 R_s^2 \sqrt{1-R_s/r}\sqrt{1-R_s/r'}} \frac{V'U +  U'V}{(\Delta U - i \epsilon)(\Delta V - i \epsilon)},
\end{equation}
\\
where $\Delta U= (U-U')$ and $\Delta V = (V-V')$. The 2-bein  moving with a freely falling detector is given by
\begin{eqnarray} \label{eq:14}
b_0^{\mu} = \left( \begin{array}{c}
  \frac{-U}{2R_s(1-\sqrt{R_s/r})} \\
  \frac{V}{2R_s(1+\sqrt{R_s/r})}
\end{array} \right),
\quad
b_1^{\mu} = \left( \begin{array}{c}
  \pm \frac{U}{2R_s(1-\sqrt{R_s/r})} \\
  \pm \frac{V}{2R_s(1+\sqrt{R_s/r})}
\end{array} \right),
\end{eqnarray}
which, after substitution in  Eq.\eqref{generaltwooold}, yields the following two-point function:
\begin{equation} \label{eq:15}
    W_{\psi,f} ^{'\text{HHI}} (x,x') = \frac{1}{16 \pi ^2 R_s^2}  \bigg[ \frac{UV'}{(1-\sqrt{R_s/r})(1+ \sqrt{R_s/r'})} +  \frac{VU'}{(1-\sqrt{R_s/r'})(1+ \sqrt{R_s/r})}  \bigg] \frac{1}{(\Delta U - i \epsilon)(\Delta V - i \epsilon)}.
\end{equation}

\subsubsection{Unruh state}
In an astrophysical context, black holes are typically formed through the collapse of massive stars, and they do not possess a white hole region. Consequently, astrophysical Schwarzschild black holes do not have radiation coming from infinity. Instead, they only emit outgoing Hawking radiation, leading to a gradual decrease in their mass. Nevertheless, due to their enormously long lifetimes, one can effectively consider the situation to be nearly time-independent. To describe such a scenario, one constructs a state, namely the Unruh state, defined only by the black hole and the universe region, and it reproduces the late-time thermal radiation \cite{Dappiaggi:2009fx}. Furthermore, the energy density and flux of outgoing radiation far away from the horizon are numerically equal in this state \cite{birrell}. One can understand the Unruh state as a squeezed state of the initial state that closely resemble the Minkowski vacuum on $ \mathcal{I}^- $. Here, one quantizes the field using positive frequency on the past horizon with respect to the null generator $\partial_U$ of $\mathcal{H}^-$, and positive frequency modes are derived with respect to the null generators $\partial_v$ on past null infinity.

The 1+1-dimensional Schwarzschild metric in terms of U and v coordinates (see Eq.~\eqref{eq:4}) is related to the Minkowski spacetime metric by a conformal factor as \( g_{\mu \nu}(x) = \Omega''^2(x) \eta_{\mu \nu}(x)\). Therefore, one can always choose a local orthogonal frame $e^a _\mu(x) = \Omega''(x) \delta ^a _\mu $ and compute the spin connection to yield the following vector parallel propagator:
\begin{equation} \label{unruhparallt}
    \Lambda ^\mu _\nu (x,x') = \frac{\Omega''(x')}{\Omega''(x)} \delta ^\mu _\nu.
\end{equation}
Substituting the above expression \eqref{unruhparallt} of the vector parallel propagator in the two-point function of the detector coupled linearly to the scalar density of a massless Dirac field, shown in Eq.\eqref{generaltwoo},  with the metric Eq.\eqref{eq:4}, we get the following two-point function:
\begin{equation} \label{confwighu}
    W_{\psi} ^{\text{Unruh}} (x,x') = -\frac{\sqrt{rr'}}{2 \pi ^2 R_s} e^{(r+r')/2R_s} e^{-(v+v')/4R_s} \frac{1}{(\Delta U - i \epsilon)(\Delta v - i \epsilon)}.
\end{equation}
Again, one can get the above Eq.\eqref{confwighu} by imposing the conformal invariance discussed in subsection 2.3.1, with knowledge of the two-point function in the Minkowski spacetime. This is the same as substituting 
\begin{eqnarray} \label{conftetru}
b_0^{\mu} = \left( \begin{array}{c}
  \frac{e^{-u/4 R_s}}{\sqrt{1-R_s/r}} \\
  \frac{e^{-u/4 R_s}}{\sqrt{1-R_s/r}}
\end{array} \right),
\quad
b_1^{\mu} = \left( \begin{array}{c}
  \mp \frac{e^{-u/4 R_s}}{\sqrt{1-R_s/r}} \\
  \pm \frac{e^{-u/4 R_s}}{\sqrt{1-R_s/r}}
\end{array} \right),
\end{eqnarray}
in Eq.\eqref{generaltwooold}  ( i.e, \( W_{\psi, c}^{'\text{Unruh}} = W_{\psi}^{\text{Unnruh}} \)).

Converting the timelike component of the above tetrad \eqref{conftetru}, which represents the four-velocity of a frame moving with tetrad, in the t-r plane by usual tetrad transformation and integrating, one gets the following trajectory:
\begin{equation}
    r_* = - u/2 - R_s e^{-u/2R_s} +C_2
\end{equation}
with proper acceleration
\begin{equation}
    a := \sqrt{a^\mu a_\mu} = \pm \frac{\cosh{(u/4R_s)}- 2R_s^2 \cosh{(u/4R_s)}/r^2 + \sinh{(u/4R_s)}}{4 R_s \sqrt{1 - R_s/r}},
\end{equation}
where $C_2$ is an integration constant. For illustrative purposes, we show the plots for the trajectory and acceleration for a particular value of \( C_2 \) in Fig. \ref{fig:0}.

The 2-bein corresponding to a static observer in the metric Eq.(\ref{eq:4}) are given by, 
\begin{eqnarray} \label{eq:16}
b_0^{\mu} = \left( \begin{array}{c}
  \frac{-U}{2R_s\sqrt{1-R_s/r}} \\
  \frac{1}{\sqrt{1-R_s/r}}
\end{array} \right),
\quad
b_1^{\mu} = \left( \begin{array}{c}
  \pm \frac{U}{2R_s\sqrt{1-R_s/r}} \\
  \pm \frac{1}{\sqrt{1-R_s/r}}
\end{array} \right).
\end{eqnarray}
Using the above 2-bein, Eq. (\ref{eq:16}), in Eq. \eqref{generaltwooold} yields the two-point function for the scalar density of the Fermionic field in the Unruh vacuum:
\begin{equation} \label{eq:17} 
    W_{\psi,s} ^{'\text{Unruh}} (x,x') = \frac{1}{8 \pi ^2 R_s \sqrt{1-R_s/r}\sqrt{1-R_s/r'}} \frac{ U +  U'}{(\Delta U - i \epsilon)(\Delta v - i \epsilon)},
\end{equation}
where $\Delta U= (U-U')$ and $\Delta v = (v-v')$. The 2-bein moving with a freely falling detector, expressed in terms of the coordinates used in metric Eq.(\ref{eq:4}), is given by
\\
\begin{eqnarray} \label{eq:18}
b_0^{\mu} = \begin{pmatrix}
  \frac{-U}{2 R_s(1-\sqrt{R_s/r})} \\
  \frac{1}{1+\sqrt{R_s/r}}
\end{pmatrix},
\quad
b_1^{\mu} = \begin{pmatrix} 
  \mp \frac{U}{2R_s(1-\sqrt{R_s/r})} \\
  \mp \frac{1}{1+\sqrt{R_s/r}}
\end{pmatrix} ,
\end{eqnarray}
which, after substitution in Eq.\eqref{generaltwooold}, yields the following two-point function:
\begin{equation} \label{eq:19}
    W_{\psi,f} ^{'\text{Unruh}} (x,x') = \frac{1}{8 \pi ^2 R_s}  \bigg[ \frac{U}{(1-\sqrt{R_s/r})(1+ \sqrt{R_s/r'})} +  \frac{U'}{(1-\sqrt{R_s/r'})(1+ \sqrt{R_s/r})}  \bigg] \frac{1}{(\Delta U - i \epsilon)(\Delta v - i \epsilon)}.
\end{equation}

\subsubsection{Boulware vacuum}
The Boulware vacuum corresponds to the vacuum state that replicates the Minkowski vacuum at both past and future null infinity. In other words, in this scenario, there is neither incoming nor outgoing radiation at infinity. Therefore, one can describe the vacuum outside compact, spherically symmetric objects like a static neutron star by Boulware vacuum \---\ given the absence of both past and future event horizons in such objects. It exhibits nonregularity on both past and future horizons. The modes corresponding to this state are defined to be the positive and negative frequency with respect to the Schwarzschild timelike Killing field, denoted as $\partial_t$. 

Again, the 1+1-dimensional Schwarzschild metric in terms of u and v coordinates (see Eq.~\eqref{eq:2}) is related to the Minkowski spacetime metric by a conformal factor as \( g_{\mu \nu}(x) = \Omega'''^2(x) \eta_{\mu \nu}(x)\). Therefore, one can again choose a local orthogonal frame $e^a _\mu(x) = \Omega'''(x) \delta ^a _\mu $ and compute the spin connection to yield the following vector parallel propagator:
\begin{equation} \label{boulparall}
    \Lambda ^\mu _\nu (x,x') = \frac{\Omega'''(x')}{\Omega'''(x)} \delta ^\mu _\nu.
\end{equation}
Substituting the above expression \eqref{boulparall} of the vector parallel propagator in the two-point function of the detector coupled linearly to the scalar density of a massless Dirac field, shown in Eq.\eqref{generaltwoo},  with the metric Eq.\eqref{eq:4}, one gets the following two-point function:
\begin{equation} \label{confwighb}
    W_{\psi} ^{\text{Boulware}} (x,x') = -\frac{\sqrt{rr'}}{2 \pi ^2 R_s} e^{(r+r')/2R_s} e^{-(v+v')/4R_s} e^{(u+u')/4R_s} \frac{1}{(\Delta u - i \epsilon)(\Delta v - i \epsilon)},
\end{equation}
which can also be obtained by assuming the conformal invariance discussed in subsection 2.3.1. This is the same as the expression obtained by substituting
\begin{eqnarray} \label{conftetrb}
b_0^{\mu} = \left( \begin{array}{c}
  \frac{1}{\sqrt{1-R_s/r}} \\
  \frac{1}{\sqrt{1-R_s/r}}
\end{array} \right),
\quad
b_1^{\mu} = \left( \begin{array}{c}
  \mp \frac{1}{\sqrt{1-R_s/r}} \\
  \pm \frac{1}{\sqrt{1-R_s/r}}
\end{array} \right),
\end{eqnarray}
in Eq.\eqref{generaltwooold} (i.e, \( W_{\psi, c}^{'\text{Boulware}} = W_{\psi, s}^{'\text{Boulware}}= W_{\psi}^{\text{Boulware}} \)). The above 2-bein corresponds to a static frame. The two-point function shown in Eq. \eqref{confwighb} can also be rewritten as
\begin{equation} \label{eq:21}
    W_{\psi} ^{\text{Boulware}} (x,x') = W_{\psi,s} ^{'\text{Boulware}} (x,x')=  -\frac{1}{2 \pi ^2 \sqrt{1-R_s/r}\sqrt{1-R_s/r'}} \frac{1}{(\Delta u - i \epsilon)(\Delta v - i \epsilon)},
\end{equation}
where $\Delta u= (u-u')$ and $\Delta v = (v-v')$. While the 2-bein moving with a freely falling detector in terms of metric Eq.(\ref{eq:2}) is given by
\begin{eqnarray}  \label{eq:22}
b_0^{\mu} = \begin{pmatrix}
  \frac{1}{1-\sqrt{R_s/r}} \\
  \frac{1}{1+\sqrt{R_s/r}}
\end{pmatrix}, 
b_1^{\mu} = \begin{pmatrix}
  \pm \frac{1}{1-\sqrt{R_s/r}} \\
  \mp \frac{1}{1+\sqrt{R_s/r}}
\end{pmatrix},
\end{eqnarray}
which, after substitution in Eq.\eqref{generaltwooold}, yields the following two-point function
\begin{equation} \label{eq:23}
    W_{\psi,f} ^{'\text{Boulware}} (x,x') = -\frac{1}{4 \pi ^2}  \bigg[ \frac{1}{(1-\sqrt{R_s/r})(1+ \sqrt{R_s/r'})} +  \frac{1}{(1-\sqrt{R_s/r'})(1+ \sqrt{R_s/r})}  \bigg] \frac{1}{(\Delta u - i \epsilon)(\Delta v - i \epsilon)}.
\end{equation}

\begin{figure*}[ht!] 
        \centering
        \includegraphics[width=0.92\textwidth]{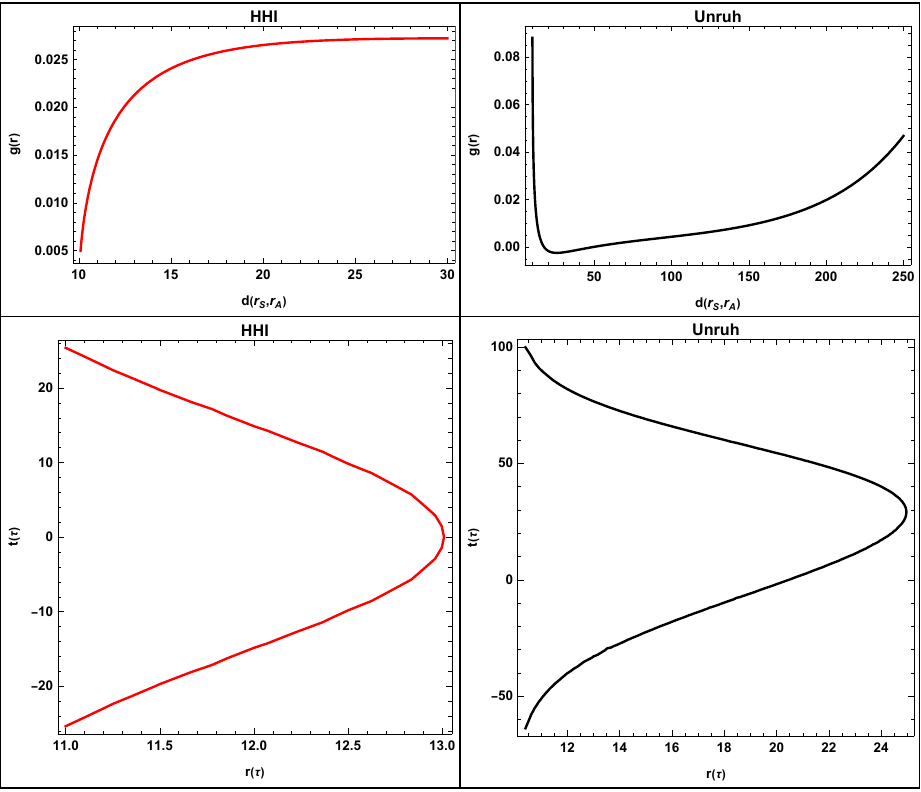}
        \captionsetup{margin=1cm, font=small}
        \caption{The top panel shows the proper acceleration of a frame where the two-point function for the interaction Hamiltonian $\hat{H} _{\text{j}}^{'\text{int}}$ shown in Eq. \eqref{eq:24old} is the same as the two-point function for $\hat{H} _{\text{j}}^{\text{int}}$ shown in Eq. \eqref{eq:24} i.e, $W_\psi^\alpha = W_{\psi,c}^{'\alpha}$. The bottom panel shows the corresponding trajectory with the integration constants $C_1$= 1 and $C_2$ = 78.}    
        \label{fig:0}
\end{figure*}

\begin{figure*}[ht!] 
        \centering
        \includegraphics[width=0.92\textwidth]{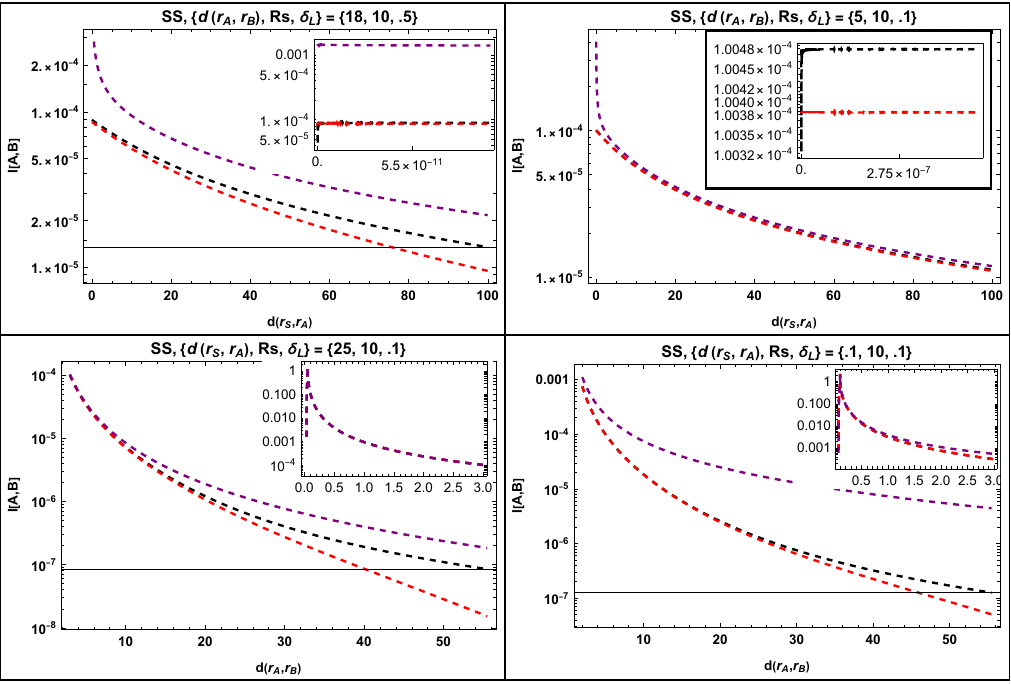}
        \captionsetup{margin=1cm, font=small}
        \caption{The above plots show the mutual information between two disjoint intervals, \([A, B]\) and \([C, D]\), in a static frame, calculated using the resolvent technique described in Section 2.4. In the plots, the black line represents the Unruh state, the red line represents the Hartle-Hawking-Israel (HHI) state, and the purple line represents the Boulware state.  We choose the intervals such that they are at a constant Schwarzschild time slice, and have a spatial extent determined by \(\delta_L\). In the top panel, we keep the proper distance between two disjoint intervals,  $d(r_A, r_B)$, constant, and vary the distance from the event horizon, $d(r_S, r_A)$. In the bottom panel, we keep the proper distance from the event horizon, $d(r_S, r_A)$, fixed and vary the proper distance between the two disjoint intervals, $d(r_A, r_B)$.}    
        \label{fig:1}
\end{figure*}

\subsection{Relative Entanglement entropy of the massless Dirac field} \label{sec:2.4}

We briefly review the calculation of the relative entanglement entropy of a massless fermionic field for disjoint intervals in the Schwarzschild spacetime using the two-point function. We shall see that it yields various interesting properties and inequalities associated with the mutual information of the field in various states.

The vacuum states defined in the last section are pure. However, if one considers a restriction to a certain region of spacetime, the reduced state will generally not be pure. The Reeh-Schlieder theorem implies that the reduced state is mixed \cite{Witten:2018zxz}. From an algebraic point of view, all von Neumann algebras associated with local observables of well-behaved QFTs are isomorphically related. However, algebras associated with local observables in different regions of spacetime can be related in a completely different manner for different QFTs. The entanglement is a kind of quantum correlation between local observables, and it is a property of the algebra of observables. However, observers restricted to certain regions of spacetime and following some specified trajectories can observe different correlations due to their different natural choices of Hilbert space decomposition. In the next section, we will discuss entanglement using localized quantum probes. The entanglement properties studied by the detector depend on several factors, including the choice of detector model. To understand what the quantum probes observe and to compare it with the existing literature on entanglement using various conformal field theory methods, in this section, we first study the relative entanglement entropy  --- a measure of total correlation --- of the field in different states using the resolvent technique \cite{Guo:2022ivd, Calabrese:2009qy, Matsuo:2021mmi}. The von Neumann/ entanglement entropy associated with a region having reduced density matrix $\rho_A = \text{Tr}_B |\psi\rangle \langle \psi| $ \---\ which is obtained by tracing out degrees of freedom outside that region   \---\ is defined by
\begin{equation} \label{eqn:27}
    S_A := - \text{Tr}(\rho_A 
    \log{\rho_A}) .
\end{equation}
The entanglement entropy of the real massless fermionic field in terms of the Wightman function, Eq.(\ref{eqn:13}), is given by\cite{Casini:2009sr}

\begin{equation}
    S = - \text{Tr} [(1-S^+) \log{(1-S^+)} + S^+ \log{S^+}].
\end{equation}
Using the integral representation of the logarithm, the above expression can be written as
\begin{equation} \label{eqn:28}
    S = - \int_{1/2}^\infty d\beta \text{Tr}\bigg[ ( \beta -1/2 ) (R(\beta) - R (-\beta)) - \frac{2 \beta}{\beta +  1/2}    \bigg],
\end{equation}
where $ R$ is the resolvent of the Wightman function, $S^+$ , and it is given by
\begin{equation}
    R = (S^+ - 1/2 + \beta)^{-1} .
\end{equation}
Using the definition of functional inverse 
\begin{equation} \label{eqn:29}
    \int d\tilde{z} R(\beta;\tilde{x},\tilde{z} )R^{-1}(\beta;\tilde{z},\tilde{y}) = \delta (\tilde{x}-\tilde{y})
\end{equation}
we get
\begin{equation} \label{eqn:30}
    \int d\tilde{z} R(\beta;\tilde{x},\tilde{z} ) [S^+(\tilde{z},\tilde{y}) + (\beta - 1/2)\delta (\tilde{z},\tilde{y}) ] =  \delta (\tilde{x}-\tilde{y}).
\end{equation}
For the case of massless Dirac field in Minkowski spacetime, the use of the correlator shown in Eq.(\ref{eqn:13}) in the above equation yields
\begin{equation}\label{eqn:31}
    \beta R(\tilde{x},\tilde{y}) - \frac{i}{2 \pi} \int \frac{R(\tilde{x},\tilde{z})}{\tilde{z}-\tilde{y}} d\tilde{z} = \delta(\tilde{x}-\tilde{y}).
\end{equation}
Assuming the region over which the integration is being performed to have n disjoint spatial intervals, ($p_i,q_i$), one gets the following solution of Eq.(\ref{eqn:31}):
\begin{equation} \label{eqn:32}
    R(\beta; \tilde{x},\tilde{y}) = (\beta ^2 -1/4)^{-1} \bigg( \beta \delta (\tilde{x}-\tilde{y} ) + \frac{i \gamma^0 \gamma^1}{2 \pi (\tilde{x}-\tilde{y})} \exp \bigg\{- \frac{i}{2\pi} \gamma^0 \gamma^1 \log{\bigg( \frac{\beta -1/2}{\beta +1/2} \bigg) } (f(\tilde{x}) - f (\tilde{y})) \bigg\}  \bigg),
\end{equation}
where
\begin{equation}
    f(\tilde{x}) = \log{ \bigg( - \frac{\prod_{i=1}^n (\tilde{x}_i - p_i) }{ \prod_{i=1}^n (\tilde{x}_i - q_i) }   \bigg) }.
\end{equation}
Substituting above resolvent into Eq.( \ref{eqn:28}) and performing integrations with $\tilde{x} \rightarrow \tilde{y} $ \cite{Guo:2022ivd, Calabrese:2009qy, Matsuo:2021mmi} one gets 
\begin{equation} \label{eqn:34}
    S = \frac{1}{3} \bigg( \sum_{i,j}  \log{|q_i - p_j|}   -\sum_{i<j}  \log{ |q_i - q_j|}   -\sum_{i<j} \log{|p_i - p_j|} - n \log{\epsilon} \bigg).
    \end{equation}
The above expression of entropy is dependent on the UV cutoff $\epsilon$, which appears because of the presence of high-energy vacuum fluctuations. However, assuming $\epsilon$ to be a constant, the relative entropy of different regions will be independent of $\epsilon$.

By repeating the above calculation for the correlator Eq.(\ref{eqn:14}) for an interval in a spacetime conformal to (1+1) dimensional Minkowski spacetime, i.e, $g_{\mu \nu}$ = $  \Omega^2 \eta_{\mu \nu}$,  one gets the following expression of entanglement entropy, in terms of null coordinates (u,v), for a massless Dirac field \cite{Matsuo:2020ypv, Calabrese:2004eu, Bianchi:2014bma} :
\begin{equation} \label{eqn:35}
    S_{[A,B]} = \frac{1}{12}  \log{ \bigg( \frac{ (v_1 - v_2)^2(u_1 - u_2)^2 \Omega^2 (v_1,u_1) \Omega^2 (v_2,u_2)}{\epsilon^2}  \bigg) },
\end{equation}
\begin{figure*}[ht!] 
        \centering
        \includegraphics[width=0.92\textwidth]{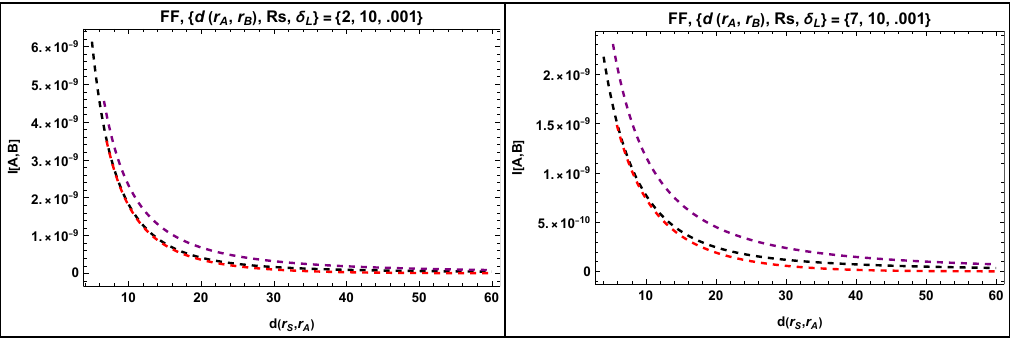}
        \captionsetup{margin=1cm, font=small}
        \caption{The above plots show the mutual information between two disjoint intervals in a freely falling frame computed using the resolvent technique discussed in section 2.4. In the plots, the black line represents the Unruh state, the red line represents the Hartle-Hawking-Israel (HHI) state, and the purple line represents the Boulware state. We keep the proper distance between two disjoint intervals at a fixed Painleve-Gulstrand time slice constant and vary the distance from the event horizon.}    
        \label{fig:2}
\end{figure*}
with ($u_1,v_1$) and ($u_2,v_2$) being null coordinates of the corners of the interval. The above expression, apart from $\epsilon$, is simply one-third of the natural logarithm of the conformal distance between two corners \footnote{The conformal distance transforms covariantly under all conformal isometries of the conformally flat spacetime \cite{conformal}.}. Using Eq.\eqref{eqn:34} again for the two disjoint regions $[A,B]$ and $ [C,D]$ one can obtain $S_{[A,B] \bigcup [C,D]}$. The substitution of the resultant expression along with Eq.\eqref{eqn:35} in the definition of mutual information 
\begin{equation}
    I := S_{[A,B]} + S_{[C,D]} - S_{[A,B] \bigcup [C,D]}
\end{equation}
we get \footnote{This is valid for a massless fermionic field of Dirac type with central charge c = 1 in 1+1 dimensions.} \cite{Casini:2004bw, PhysRevD.79.024015, Casini:2008wt, Calabrese:2009ez, PhysRevD.106.086019,  Headrick:2019eth},
\begin{eqnarray} \label{mutualmain}
I = \frac{1}{12} \log{ \left( \frac{ (v_1 - v_3)^2(u_1 - u_3)^2 (v_2 - v_4)^2(u_2 - u_4)^2}{ (v_3 - v_2)^2(u_3 - u_2)^2 (v_1 - v_4)^2(u_1 - u_4)^2} \right) }.
\end{eqnarray}

Here, $(u_1, v_1)$ and $(u_2, v_2)$ are the null coordinates of the corners of the spacelike interval [A, B], and $(u_3, v_3)$ and $(u_4, v_4)$ are those of the spacelike interval [C, D]. All corners are ordered by increasing/decreasing spatial coordinates. The above expression of mutual information between two disjoint intervals is Weyl invariant. In particular, it remains invariant under any single transformation or any combination of the following transformations:
\begin{enumerate}
    \item $u_1 \rightarrow 1/u_1$, $u_2 \rightarrow 1/u_2$, $u_3 \rightarrow 1/u_3$, $u_4 \rightarrow 1/u_4$ \\
    \item $v_1 \rightarrow 1/v_1$, $v_2 \rightarrow 1/v_2$, $v_3 \rightarrow 1/v_3$, $v_4 \rightarrow 1/v_4$ \\
    \item $u_1 \rightarrow 1/u_3$, $u_2 \rightarrow 1/u_4$, $u_3 \rightarrow 1/u_1$, $u_4 \rightarrow 1/u_1$ \\
   \item $v_1 \rightarrow 1/v_3$, $v_2 \rightarrow 1/v_4$, $v_3 \rightarrow 1/v_1$, $v_4 \rightarrow 1/v_1$ \\
   \item $u_1 \rightarrow 1/v_1$, $u_2 \rightarrow 1/v_2$, $u_3 \rightarrow 1/v_3$, $u_4 \rightarrow 1/v_4$, $v_1 \rightarrow 1/u_1$, $v_2 \rightarrow 1/u_2$, $v_3 \rightarrow 1/u_3$, $v_4 \rightarrow 1/u_4$
\end{enumerate}

One can observe from Eq.\eqref{mutualmain} that the mutual information \( I \) reaches its maximum when the separation between \((u_1, v_1)\) and \((u_4, v_4)\), or between \((u_2, v_2)\) and \((u_3, v_3)\), is null. Conversely, it reaches its minimum when the separation between \((u_1, v_1)\) and \((u_3, v_3)\), or between \((u_2, v_2)\) and \((u_4, v_4)\), is null. However, since we take the intervals to belong to non-intersecting causal domains, these conditions are not satisfied, leading instead to a finite and nonnegative mutual information\footnote{At this point, it is important to recall that the mutual information of a bipartite system is always non-negative. However, the tripartite mutual information of two extended intervals can be negative \cite{Casini:2008wt}.}. Nevertheless, by keeping both intervals non-intersecting while bringing \((u_2, v_2)\) and \((u_3, v_3)\) close to null, one can observe a peak in the mutual information plots (see Fig.\ref{fig:1} and Fig.\ref{fig:5}). This point acts as a phase transition point of the entanglement. One can refer to \cite{Molina-Vilaplana:2011ydi} for a discussion of the phase transition of the mutual transition by varying the separation between adjacent intervals. Furthermore, the relation of the mutual information with entanglement negativity between adjacent disjoint intervals has been discussed in \cite{Calabrese:2012nk, Kudler-Flam:2018qjo}.

The properties of mutual information outlined above are applicable to all three vacua discussed in the previous section. To study different other properties, we first write down the mutual information explicitly.
Using Eq.\eqref{mutualmain}, one gets the following expressions of the mutual information for respective vacua :
\begin{align} 
I_{\text{HHI}} &= \frac{1}{12} \log{ \left( \frac{ (V_1 - V_3)^2(U_1 - U_3)^2 (V_2 - V_4)^2(U_2 - U_4)^2}{ (V_3 - V_2)^2(U_3 - U_2)^2 (V_1 - V_4)^2(U_1 - U_4)^2} \right) } ,  \label{mihhi}\\
I_{\text{Unruh}} &= \frac{1}{12} \log{ \left( \frac{ (v_1 - v_3)^2(U_1 - U_3)^2 (v_2 - v_4)^2(U_2 - U_4)^2}{ (v_3 - v_2)^2(U_3 - U_2)^2 (v_1 - v_4)^2(U_1 - U_4)^2} \right) }, \label{miunruh} \\
I_{\text{Boulware}} &= \frac{1}{12} \log{ \left( \frac{ (v_1 - v_3)^2(u_1 - u_3)^2 (v_2 - v_4)^2(u_2 - u_4)^2}{ (v_3 - v_2)^2(u_3 - u_2)^2 (v_1 - v_4)^2(u_1 - u_4)^2} \right) }.  \label{miboul}
\end{align}
We have used Eq.~\eqref{eqn:13} to derive the expression for the entanglement entropy in Eq.~\eqref{eqn:34}, which holds on a constant time slice. Therefore, the expressions given in Eqs.~\eqref{mihhi}--\eqref{miboul} for \( I_{\text{HHI}} \), \( I_{\text{Unruh}} \), and \( I_{\text{Boulware}} \) are defined for a constant Schwarzschild time slice. These expressions, can be interpreted by rewriting \( U \) and \( V \) in terms of \( \sinh \) and \( \cosh \), and noting that the Hartle-Hawking-Israel state is thermal in both the outgoing and ingoing modes, while the Unruh state is thermal in the outgoing modes and in the vacuum state for the ingoing modes. 

%Assuming the thermal behavior of HHI and Unruh states remains valid by replacing the Schwarzschild time by the Painleve Gulstrand time, $t_{PG}$, one can get the expression of the mutual information on a constant $t_{PG}$ slice by replacing $U_s,V_s,u_s,v_s $ in Eqs.\eqref{mihhi}--\eqref{miboul} by $ U' = -2 R_s e^{-u' /2R_s}, \quad V = 2 R_s e^{v' / 2R_s}$, with $u'= t_{PG}-r_*$,  $v'= t_{PG}+r_*$.

The Kruskal coordinates \( U \) and \( V \) are given by:
$ U = -2 R_s e^{-u /2R_s}, \quad V = 2 R_s e^{v / 2R_s}$, with $u= t-r_*$,  $v= t+r_*$. Let us introduce the dimensionless variables \(\eta_s = v_s/2R_s\) and \(\xi_s = u_s/2R_s\). The difference between the Boulware and Hartle-Hawking-Israel mutual information is then expressed as
\begin{equation}
I_{\text{Boulware}} - I_{\text{HHI}} = \frac{1}{12} \log \left( \frac{h(\xi_4)}{h(\xi_3)} \frac{g(\eta_4)}{g(\eta_3)} \right),
\end{equation}
\\
where the functions \(g(\eta_s)\) and \(h(\xi_s)\) are defined in Appendix \ref{Appendix E}. Taking $\xi_1 > \xi_2 > \xi_3 > \xi_4$ (or, $\xi_1 < \xi_2 < \xi_3 < \xi_4$ ), the monotonicity of the function $h(\xi_s)$, under the assumptions given in Appendix \ref{Appendix E}, implies $h(\xi_4) > h(\xi_3)$. Furthermore, taking $\eta_1 < \eta_2 < \eta_3 < \eta_4$ (or,  $\eta_1 > \eta_2 > \eta_3 > \eta_4$), the monotonicity of the function $g(\eta_s)$, under the considerable assumptions in Appendix \ref{Appendix E}, implies $g(\eta_4) > g(\eta_3)$. Therefore, $ I_{\text{Boulware}} >  I_{\text{HHI}}$. Repeating the procedure outlined above, one can see that 
\begin{equation}
    I_{\text{Boulware}} > I_{\text{Unruh}} > I_{\text{HHI}}
\end{equation}
for sufficiently separated regions A and B. For illustration purposes, we show the plots for the mutual information in different vacua from the perspective of a static observer in Fig.~\ref{fig:1} and, from the perspective of a freely falling observer in Fig.\ref{fig:2}. Therefore, the total correlation between the two disjoint intervals is minimal for the HHI vacuum compared to the other two vacua, while it is maximal for the Boulware vacuum. This difference can be attributed to the presence of Hawking radiation in the HHI and Unruh states, as opposed to only vacuum polarization in the Boulware state. It suggests that the presence of Hawking radiation for a static observer reduces the total correlation. We will defer a detailed analysis of the shapes of the plots in Fig.~\ref{fig:1} and Fig.~\ref{fig:2} until Section \ref{section4}, where we will also explore the mutual information derived from entanglement harvesting using detectors. 

Having described various vacua and the corresponding entanglement properties, we now proceed to the protocol of entanglement harvesting in the next section.

\begin{figure*}[ht!] 
        \centering
        \includegraphics[width=0.92\textwidth]{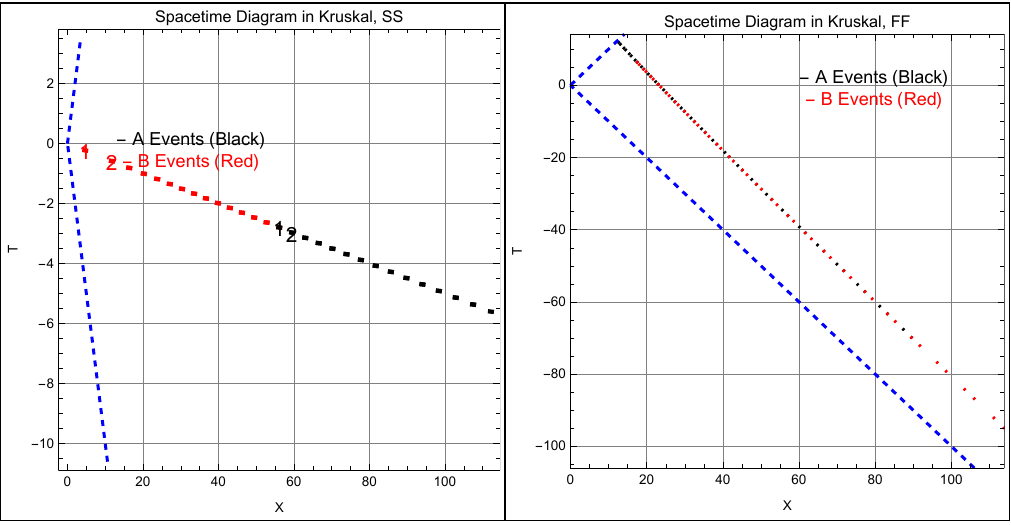}
        \captionsetup{margin=1cm, font=small}
        \caption{The left panel shows the disjoint intervals for static observers, while the right panel shows those for freely falling observers. The blue dashed lines represent the event horizon. Correlation measures are computed between the disjoint intervals: red 1 with black 1, red 2 with black 2, and so on.}    
        \label{fig:interval}
\end{figure*}

\section{Introducing Unruh-DeWitt detector formalism} \label{section3}
Entanglement harvesting is a protocol where one couples certain quantum probes, initially prepared in a joint separable state, to a shared environment. The probes interact locally with the common environment, and after a certain time, we observe that the probes become entangled with each other due to the entanglement inherent in the environment. This protocol can extract the entanglement within the field, even when we do not allow direct communication between the probes \cite{reznik, Alejandro}. The forecasts made by widely employed particle detectors with spatial smearing, coupled to quantum fields, typically lack general covariance beyond the pointlike limit. The degree of violation of covariance is contingent upon various factors, including the state of the detectors-field system, the configuration and motion status of the detectors, and the characteristics of the spacetime geometry \cite{martin2}. The absence of covariance becomes evident through an ambiguity in the time-ordering operation. Additionally, the quantum field theories lack well-defined position observables. Nevertheless, the Unruh-DeWitt detector models play a crucial role as a theoretical framework for extracting localized spatiotemporal information from quantum fields. Here, we use a point like Unruh-DeWitt detector (UDW), a two-level system, as a useful theoretical tool for extracting entanglement from the field in Schwarzschild spacetime that admits a Wightman function. 

Let the interaction between the field and the $j$-th detector be governed by the following interaction picture Hamiltonian, which is the closest spinor field equivalent to the scalar field UDWs \cite{takagi,jorma, mokhtar, Wu:2023ych, Mandl:1985bg}:
\begin{equation} \label{eq:24}
   \hat{H} _{\text{j}}^{\text{int}}(\tau_j)  = \lambda _j \chi_j (\tau _j) \hat{\mu}(\tau_j) :\bar{\psi}(x (\tau _j))\psi(x(\tau_j)): .
\end{equation}
Here $\tau_j$ is the proper time in the frame of $j^{th}$ detector, $\lambda_j$ is a dimensionless coupling constant, $\chi_j(\tau _j)$ represents the switching function and $\hat{\mu}(\tau_j)$ is the monopole coupling of $j^{th}$ detector. Since the Gaussian switching is widely used in the literature on scalar fields \cite{tjoaman, gytm}, we have chosen to employ the same Gaussian switching profile, denoted as $\chi_j(\tau_j) = \exp \left(-(\tau_j - \tau_{j,0})^2/2 \sigma_j ^2 \right)$. One can restore the most common form of switching function $ \exp \left(-(\tau_j - \tau_{j,0})^2/ \tilde{ \sigma_j} ^2 \right)$ by rescaling $\sigma \rightarrow \tilde{\sigma}/\sqrt{2} $. Since the Gaussian switching function dies down rapidly, for the sake of numerical simplicity, we do our numerical integrations in the interval $[-5\sigma_j + \tau_{j,0}, 5\sigma_j + \tau_{j,0}]$ \cite{tjoaman, gytm}.  

The above monopole coupling shown in Eq.\eqref{eq:24} can be viewed as a simplified version of the Hamiltonian describing the coupling of an electron with a photon, which is proportional to $A_\mu \bar{\psi} \gamma ^\mu \psi$ \cite{Hummer}. The calculation of two point function in Appendix [\ref{Appendix A}], for the interaction Hamiltonian $ \hat{H} _{\text{j}}^{\text{int}}$ shown in Eq.\eqref{eq:24}, involves a spin parallel propagator $U(x,x')$.  Since we are working in 1+1 dimensions, the local Lorentz group is abelian and its spinor representation is one-dimensional. So, the spin connection reduces to a single one-form, and we can decompose the spinor parallel propagator as \( U(x', x) = \bar{A}(x') A(x), \quad U(x, x') =  \bar{A}(x) A(x') \) at every spacetime point. We then define a transformed spinor field as \( \psi'(x) = A(x) \psi(x), \quad \bar{\psi}'(x) = \bar{A}(x) \bar{\psi}(x). \) In subsequent subsections, we will also discuss the UDWs coupled to the transformed field with the following interaction Hamiltonian: 
\begin{equation} \label{eq:24old}
   \hat{H} _{\text{j}}^{'\text{int}}(\tau_j)  = \lambda _j \chi_j (\tau _j) \hat{\mu}(\tau_j) :\bar{\psi'}(x (\tau _j))\psi'(x(\tau_j)): .
\end{equation}
The spin parallel propagator \( U(x,x') \) transforms under a local Lorentz transformation as \( U(x,x') \rightarrow D[L(x)]\, U(x,x')\, D[L(x')]^{-1} \), where $ D[L(x)]$ is the spinor representation of the Lorentz transformation at point $x$. Factorising \( U(x,x') \) as  \( U(x,x') = \, \bar{A}(x) A(x')
\) requires the combination \( \, \bar{A}(x) A(x') \) also to transform as 
\(\, \bar{A}(x)  A(x') \rightarrow D[L(x)]\,\, \bar{A}(x)  A(x')\, D[L(x')]^{-1}. \) However, individually \( A(x') \) and \( \bar{A}(x) \) do not transform as  \(A(x') \rightarrow D[L(x')]\, A(x')\, D[L(x')]^{-1}, \quad 
\bar{A}(x) \rightarrow D[L(x)]\, \bar{A}(x)\, D[L(x)]^{-1}\)
since this does not yield the required transformation property for \( U(x,x') \). So, this makes the two-point function \( W_{\psi,b}^{'\alpha} (x,x') \) shown in Eq.~\eqref{generaltwooold}, and derived in Appendix~\ref{Appendix A}, dependent on the choice of local Lorentz frame through its dependence on the 2-bein.

The following discussions for $\hat{H}$ and $\hat{H'}$ will be similar, so we will not repeat them for both. The time evolution of $\hat{\mu}(\tau_j)$ in the interaction picture, governed by the unperturbed Hamiltonian, is expressed as 
\begin{equation} \label{eq:25}
  \hat{\mu}(\tau_j)   = \sigma _j ^{+} e^{i \Omega _j \tau_j} + \sigma _j ^{-} e^{- i \Omega _j \tau_j}.
\end{equation}
The operators $\sigma ^{\pm}$ in the above expression represent the SU(2) ladder operators, and $\Omega _j$ denotes the energy gap between the detector's ground state $|0 \rangle $ and excited state $|1 \rangle $. The ladder operators acts as: $\sigma ^ +$ $|0 \rangle$ = $|1 \rangle$, $\sigma ^ +$ $|1 \rangle$ = 0, and $\sigma ^ -$ $|1 \rangle$ = $|0 \rangle$, $\sigma ^ -$ $|0 \rangle $ = 0. The proper time in the frame of each detector is different. Therefore, we use coordinates based on the proper time of a freely falling test particle from infinity, known as the Painlevé-Gullstrand coordinate system, and relate all other times to Painlevé-Gullstrand time, denoted as $t_{PG}$. We set the origin of our  Painlevé-Gullstrand time such that the detector starts at rest from infinity at $t_{PG} = -\infty$ and reaches the singularity at $r=0$ at $t_{PG} = 0$.

The total interaction Hamiltonian for the system of two detectors, A and B, coupled to the background Dirac field, can be written as
\begin{equation} \label{eq:26}
    \hat{H}_{\text{tot}}^{\text{int}}(t_{PG}) = \frac{d\tau_A}{dt_{PG}} \hat{H} _A ^{\text{int}} (\tau_A(t_{PG})) \otimes \mathds{1} _B +  \mathds{1} _A \otimes \frac{d\tau_B}{dt_{PG}} \hat{H} _B ^{\text{int}} (\tau_B(t_{PG})) .
\end{equation}
\\
Here $\hat{H} _A ^{\text{int}}$ and $\hat{H} _B ^{\text{int}}$ represent interaction Hamiltonians of detectors A and B, respectively and we have employed the fact that both the reparametrized and initial Hamiltonians satisfy the Schrödinger equation, which can be expressed as follows: $\hat{H}(t_{PG}) |\psi \rangle = i \frac{d}{dt} |\psi \rangle$ and $\hat{H}(\tau) |\psi \rangle = i \frac{d}{d\tau} |\psi \rangle$. For simplicity, we consider the detectors to be identical; therefore, we have $\Omega_A = \Omega_B = \Omega$. We also assume that both the field and the detector are initially in their respective ground states at the moment when we switch on the detector. Therefore, we can write the initial state as 
\begin{equation} \label{eq:27}
    \hat{\rho}_{AB (t_{PG} =0)} = |0_A\rangle \langle 0_A|\otimes |0_B\rangle \langle 0_B| \otimes|0_\alpha \rangle \langle 0_\alpha| .
\end{equation}
The evolution of the state can be described by the time evolution operator $\hat{U}_I = \mathcal{T}_{t_{PG}} \exp{\left(-i \int_{\mathds{R}} dt_{PG} \hat{H}_{\text{tot}}^{\text{int}}(t_{PG})\right)}$, where $\mathcal{T}_{t_{PG}}$ represents the time ordering symbol. Taking the coupling constant $\lambda$ to be small and tracing over the field degrees of freedom, one obtains the following reduced density matrix in the standard basis (i.e., $|00\rangle, |01\rangle, |10\rangle, |11\rangle $) to the lowest order in $\lambda$ \cite{reznik,gytm}:
\begin{equation} \label{eq:28}
\rho_{AB} =
    \left( 
          \begin{array} {cccc} 
          1 - \mathcal{L}_{AA}- \mathcal{L}_{BB}  & 0 & 0 &  \mathcal{M^*}
     \\ 
          0 & \mathcal{L}_{BB} & \mathcal{L}_{BA} & 0 
     \\ 
          0 & \mathcal{L}_{AB} & \mathcal{L}_{AA} & 0 
     \\ 
         \mathcal{M} & 0 & 0 & 0
          \end{array}
     \right) + \mathcal{O} (\lambda ^4). 
\end{equation} 
The matrix elements are specified as follows:
\begin{equation} \label{eq:29}
    \mathcal{L}_{ij} = \lambda ^2 \int _{-\infty} ^{\infty} d\tau_i \int _{-\infty} ^{\infty} d\tau_j ^{'} \chi _i (\tau_i) \chi _j (\tau_j ^{'} ) e^{-i \Omega (\tau_i- \tau_j ^{'})} W_\psi ^{\alpha} (x_i(\tau_i),x_j(\tau ^{'}_j)),
\end{equation}

\begin{multline} \label{eq:30}
      \mathcal{M} = -  \lambda ^2 \int _{-\infty} ^{\infty} d\tau_A \int _{-\infty} ^{\infty} d\tau_B  \chi _A (\tau_A) \chi _B (\tau_B  ) e^{i \Omega (\tau_A + \tau_B )} \bigg[ \Theta(t_{PG}(\tau_A)- t_{PG}(\tau_B)) W_\psi ^\alpha (x_A(\tau_A),x_B(\tau _B)) + \\ \Theta(t_{PG}(\tau_B)- t_{PG}(\tau_A))  W_\psi ^\alpha (x_B(\tau_B),x_A(\tau _A)) \bigg].
\end{multline}
\\
Here, `i' and `j' represent detectors `$A$' or `$B$,' $\Theta(..)$ represents the Heaviside step function, and $W^\alpha_{\psi}(..)$ denotes the pullback of the two-point functions discussed in section 2.3 along the detector's trajectory. We note that the local contributions, i.e., $i=j$ in Eq.(\ref{eq:29}), depend on the pullback of the two-point function along the detector trajectory. In contrast, Eq. (\ref{eq:30}), which governs the nonlocal contribution, depends on the corresponding Feynman propagator, represented by the term in parentheses.
\begin{figure*}[ht!] 
        \centering
        \includegraphics[width=0.92\textwidth]{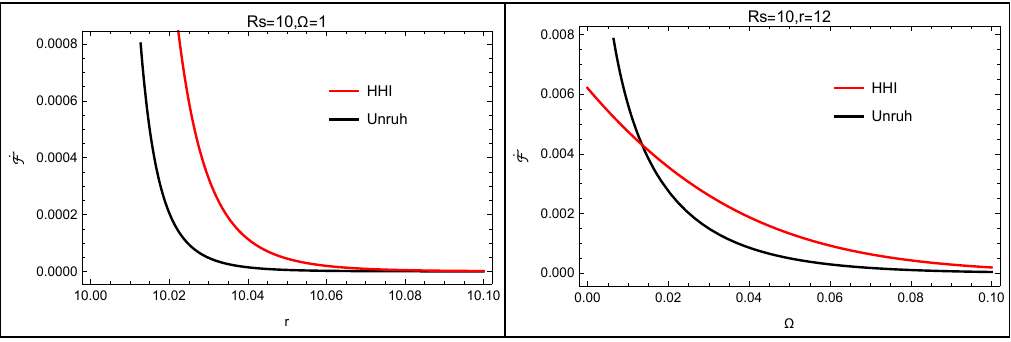}
        \captionsetup{margin=1cm, font=small}
        \caption{The right plot illustrates the transition rate of a Unruh-DeWitt detector held static in both HHI and Unruh vacua, varying with the energy gap of the detector. Meanwhile, the left plot displays the transition rate as a function of the Schwarzschild radial coordinate. }    
        \label{fig:3}
\end{figure*} 
\newpage

\subsection{Transition rate}

As a preliminary to entanglement harvesting, in this subsection, we first analyze the transition rate of the detector for different states.

The elements $\mathcal{L}_{AA}$ and $\mathcal{L}_{BB}$ in the density matrix $\rho_{AB}$ represent the standard expressions for the detector responses of detectors $A$ and $B$, respectively, along a given trajectory. In contrast, the nonlocal term $\mathcal{M}$ contains the information about entanglement between two detectors. In order to understand what a static detector detects, we define the response rate to linear order in perturbation theory as \cite{birrell}
\begin{equation} \label{eq:31}
    \dot{\mathcal{F}} = \int _{-\infty} ^{\infty} d\Delta \tau e^{-i \Omega \Delta \tau}   W_\psi ^\alpha (x(\tau),x(\tau ^{'} )) ,
\end{equation}
where we have switched on the detector smoothly for an infinite time \footnote{We have dropped a factor that depends only on the structure of the detector.}. The above expression is just the Fourier transform of the pullback of the two-point function along the detector's trajectory. For a static detector, due to the timelike Killing symmetry, the two-point functions discussed in section 2.3 depend only on the proper time interval $\Delta \tau = \tau - \tau ^{'} $ and the Schwarzschild radial coordinate `r.' Using the two-point function given in Eq. \eqref{eq:21}, it can be seen that the transition rate of a static detector with a sufficiently large energy gap, and kept initially in the ground state, in the Boulware vacuum is 0. In contrast, the transition rate of a static detector in the Hartle–Hawking state, for both two-point functions $W_{\psi,s}^{'\text{HHI}}$ defined in \eqref{eq:13} and $W_{\psi,c}^{'\text{HHI}}$ = $W_{\psi}^{\text{HHI}}$ defined in \eqref{confwighh}, is given by
\begin{equation} \label{eq:32}
    \dot{\mathcal{F}} (\Omega) =  \frac{\Omega}{\pi (e^ {4 \pi \kappa R_S \Omega } - 1)}
\end{equation}
(see Appendix \ref{Appendix B} for detailed calculations), which describes a thermal bath of a bosonic field \cite{jorma}. The factor $\pi$ in the above expression would not be present if we express it in terms of frequency rather than angular frequency. It's worth emphasizing that this transition rate for spinor fields in (1+1) dimensions is twice that of Hawking radiation associated with a massless scalar field in (3+1) dimensions, as we are working with a spin-1/2 field \cite{birrell}. The computation of the transition rate of the static detector in the Unruh vacuum is detailed in Appendix \ref{Appendix C }, and the result is
\begin{align} \label{eq:33}
    \dot{\mathcal{F}} & = \frac{1}{2}\bigg[ 2 \times \frac{1}{4 \pi ^2 R_S \kappa} \log { \frac{e^ {4 \pi \kappa R_S \Omega }}{(e^ {4 \pi \kappa R_S \Omega } \pm 1)}} \bigg]\\
    & =  \frac{1}{4 \pi ^2 R_S \kappa } \bigg[ 
    \frac{1}{e^ {4 \pi \kappa R_S \Omega } \pm 1} -\frac{1}{2}  \bigg(  \frac{1}{e^ {4 \pi \kappa R_S \Omega } \pm 1}  \bigg) ^2  + \frac{1}{3}  \bigg(  \frac{1}{e^ {4 \pi \kappa R_S \Omega } \pm 1}  \bigg) ^3 + ......  \bigg]. \label{eq:34}
\end{align}
\\
Here, the plus sign corresponds to the two-point function in the conformal frame tetrad, $W_{\psi,c}^{'\text{Unruh}}$ = $W_{\psi}^{\text{Unruh}}$, as given in Eq.~\eqref{confwighu}, while the minus sign corresponds to the two-point function in the static frame tetrad, $W_{\psi,s}^{'\text{Unruh}}$, presented in Eq.~\eqref{eq:17}. We write Eq. (\ref{eq:33}) in this form to remind that there is only an outgoing flux of particles, and we are dealing with the spin-1/2 Dirac field. Remarkably, this is not proportional to the number density of Fermions/Bosons at frequency $\Omega$ in a thermal bath. Nevertheless, one can understand it by looking at the Helmholtz free energy of Fermions/Bosons in length $L$ at temperature $T$ \cite{kapusta}, which is given by
\begin{equation} \label{eq:35}
    \mathcal{H} =  2 L \int _0 ^ \infty \frac{d\Omega}{2 \pi} \bigg[ - \Omega + T \log{ \bigg( \frac{e^{\Omega/k_B T}}{e^{\Omega/k_B T} \pm 1}}  \bigg) \bigg].
\end{equation}
Here, the first term in Eq.(\ref{eq:35}) represents the zero-point energy. Hence, the transition rate of a static detector far from the black hole in the Unruh state, Eq.(\ref{eq:33}), is proportional to the Helmholtz free energy density of Fermions/Bosons at a temperature of $1/(4\pi\kappa r_S)$. This contrasts with the findings in \cite{mokhtar}, where it was observed that the response of UDW for Fermions in Minkowski spacetime with a moving wall boundary exhibits Helmholtz free energy density with Fermionic statistics. Here, the difference in statistics arises from the dependence on $b_\alpha^\mu(x) b_\beta^\nu(x')$ in the two-point functions for different tetrads in the Unruh vacuum. Due to the dependence on $b_\alpha^\mu(x) b_\beta^\nu(x')$ if one doesn't restrict the 2-bein to move with the detector, one would obtain Fermionic statistics using the conformal 2-bein. Thus the nature of the statistics is dependent on the 2-bein chosen. We can further compare the transition rates of a static detector coupled to both types of interaction Hamiltonians, namely, the one associated with the tetrad moving with the detector and the other associated with the conformal symmetry-preserving frame, in both the Unruh state and the HHI state. The corresponding plots are displayed in Fig.~\ref{fig:3}, which clearly show that the transition rate in both of these vacua decreases as we move away from the horizon. and in the Unruh state, the magnitude for high frequencies is consistently lower than that in the HHI vacuum. However, the transition rate at low frequencies is greater for the Unruh state.

\subsection{Presence of anti-Hawking effect}
The expressions for the transition rate of the detector in HHI and Unruh states, obtained in Eqs.~(\ref{eq:32}), (\ref{eq:33}), and further illustrated in Fig.~\ref{fig:3},  show that the transition rate of a static UDW detector increases as the distance from the horizon decreases. This is consistent with the fact that the KMS temperature for static observers, $T_{KMS} = \left(64 \pi^2 M^2 (1 - 2GM/r)\right)^{-1/2}$, also increases with a decrease in $r$. However, the detector can cool with an increase in temperature, an effect called the anti-Hawking phenomenon \cite{antihawking, antiunruh}. The anti-Hawking phenomenon can be quantified by the negative sign of the partial derivative of transition probability (or the excitation to de-excitation ratio) with respect to KMS temperature.  The effective temperature function for the radially infalling observer, with velocity -$\sqrt{2M/r}$ and zero proper acceleration at `r' , is given by \cite{Barbado:2016nfy, Barbado:2012pt, Barbado:2011dx, Chakraborty:2015nwa, Smerlak:2013sga}  
\begin{equation}  \label{teffdef}
    T_{\text{eff}} = \frac{1}{1 - \sqrt{ R_s / r}} \bigg( \frac{1}{2 R_s} - \frac{R_s}{2 r^2} \bigg),  
\end{equation}  
which is the sum of the Hawking and local Unruh components, namely  
\[
 \frac{1}{2 R_s (1 -\sqrt{ R_s / r})} \quad \text{and} \quad -\frac{R_s}{2 r^2 (1 -\sqrt{ R_s / r})}.
\]
The temperature perceived by the observer is given by \( T_{\text{eff}} / 2\pi \). Therefore, a negative sign of \(\partial L_{ii} / \partial T_{\text{eff}}\) should also indicate the weak anti-Hawking effect. Since the transition probability of the detector is a competing term in the entanglement measures, as we shall discuss below in section \ref{section4}, here, we first investigate the anti-hawking effect in the present subsection.

Substituting the two-point functions corresponding to the static frame tetrad, Eqs.\eqref{eq:13} and \eqref{eq:17} discussed in the previous section, in Eq.\eqref{eq:29} for $\mathcal{L}_{ij}$ and using the saddle point approximation (appendix \ref{Appendix D}) one obtains the following transition probability of a static detector
\begin{figure*}[ht!]
        \centering
        \includegraphics[width=0.92\textwidth]{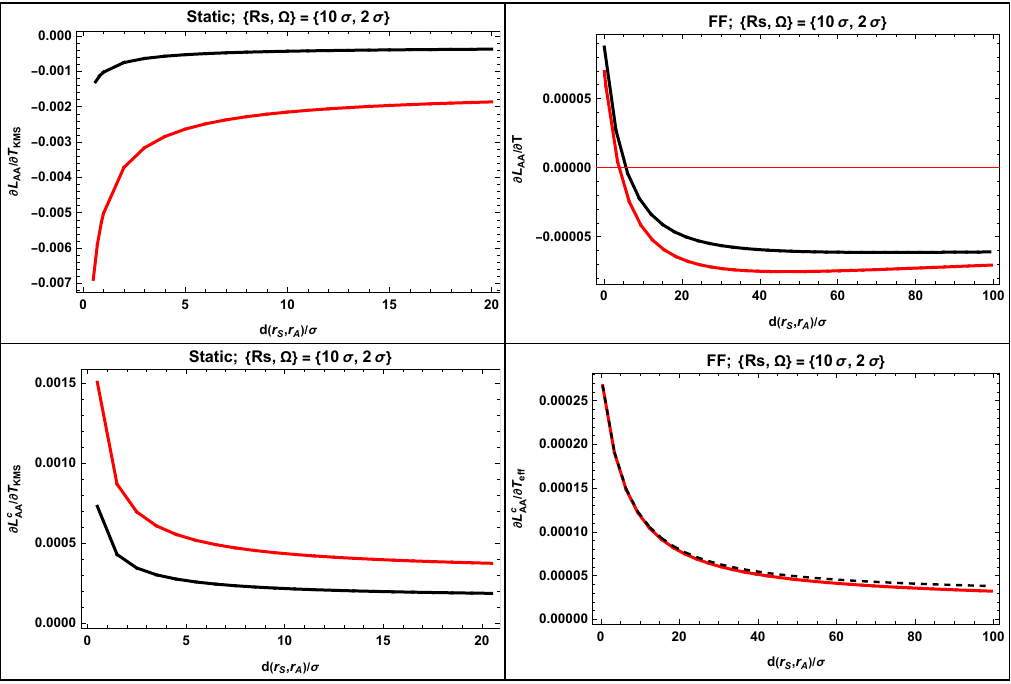}
        \captionsetup{margin=1cm, font=small}
        \caption{The above figures show the numerical plots of the partial derivative of the transition probability with respect to the KMS temperature as a function of the proper distance from the event horizon on a constant Painleve-Gullstrand time slice (x-axis). In the plots, the black line represents the Unruh state and the red line represents the Hartle-Hawking-Israel (HHI) state. The top panels display the results for the 2-bein co-moving with the detector coupled with the interaction Hamiltonian $\hat{H} _{\text{j}}^{'\text{int}}$ shown in Eq.~\eqref{eq:24old}, while the bottom panels show the results for the interaction Hamiltonian $\hat{H} _{\text{j}}^{\text{int}}$ shown in Eq.~\eqref{eq:24}. The UDW in the left panel plots are static while freely falling in the right panels.}  
         \label{fig:4}
\end{figure*}
  \begin{equation} \label{eq:36}
    L_{AA,s}^{\text{HHI}} \approx \frac{\lambda ^2 \sigma ^2}{16 \pi R_s^2 (1 - R_s/r)} e^{-\Omega ^2 \sigma ^2 } \frac{\cos{\frac{\Omega \sigma ^2}{
   R_s \sqrt{(1 - R_s/r)}} }}{\sin^2{\bigg(\frac{\Omega \sigma ^2}{
   2 R_s \sqrt{(1 - R_s/r)}}\bigg)}}
 \end{equation}
 and
  \begin{equation} \label{eq:37}
    L_{AA,s}^{\text{Unruh}} \approx \frac{\lambda ^2}{8 \pi R_s \Omega \sqrt{(1 - R_s/r)}} e^{-\Omega ^2 \sigma ^2} \cot{\frac{\Omega \sigma ^2}{
  2 R_s \sqrt{(1 - R_s/r)}} }.
 \end{equation}
Whereas, the two-point functions corresponding to the conformal symmetry preserving frame tetrad, $W_{\psi,c}^{'\text{HHI}}$ = $W_{\psi}^{\text{HHI}}$ shown in Eq.\eqref{confwighh}, and $W_{\psi,c}^{'\text{Unruh}}$ = $W_{\psi}^{\text{Unruh}}$ shown in Eqs.\eqref{confwighu}, yield the following transition probability 
 \begin{equation} \label{eqc:36}
    L_{AA,c}^{\text{HHI}} \approx \frac{\lambda ^2 \sigma ^2}{16 \pi R_s^2 (1 - R_s/r)} e^{-\Omega ^2 \sigma ^2 } \cosec^2{\bigg(\frac{\Omega \sigma ^2}{
   2 R_s \sqrt{(1 - R_s/r)}}\bigg)}
 \end{equation}
 and
  \begin{equation} \label{eqc:37}
    L_{AA,c}^{\text{Unruh}} \approx \frac{\lambda ^2}{8 \pi R_s \Omega \sqrt{(1 - R_s/r)}} e^{-\Omega ^2 \sigma ^2 } \cosec{\frac{\Omega \sigma ^2}{
  2 R_s \sqrt{(1 - R_s/r)}} }.
 \end{equation}
Here, the superscript represents the corresponding state; subscript c stands for conformal, and $\Omega \sigma ^2/ 2 \kappa R_s <\pi$ has been assumed to make sure that the contribution from the residue term vanishes \cite{saddle}.  Furthermore,  we have also assumed the sign of $\Omega$ to be positive in the above expressions. In the other case $\Omega < 0$, one would have to shift the contour in the opposite direction, which will force the contour to cross one pole even in the approximation  $\Omega \sigma ^2/ 2 \kappa R_s <\pi$, and will bring an exponential factor obeying the detailed balance form of KMS thermality in the case of the HHI state.
 
The form of Eqs. (\ref{eq:36})-(\ref{eqc:37}) for the transition probabilities, apart from a factor of cosine in the numerator, is similar to what has been found in \cite{saddle} for the scalar field in various cases. From the transition probability expressions (\ref{eq:36})-(\ref{eqc:37}), one can further compute \(\partial L_{ii}/\partial T_{KMS}\), which determines the presence of the weak anti-Hawking effect. One can also do the same by computing the transition probability integrations numerically without using the saddle point approximation. We compute \(\partial L_{ii}/\partial T_{KMS}\) numerically \footnote{Here and throughout the paper, we employ the numerical integration techniques described in \cite{tjoaman}, ensuring stable integration by setting the MinRecursion value to 3 while retaining other parameters at their default settings.} without using saddle point approximation and plot the graphs in Fig.~\ref{fig:4}. We find the plots obtained using the numerical integrations and the saddle point approximations in (\ref{eq:36})-(\ref{eqc:37}) are consistent with each other\footnote{We note that, for the numerical evaluation of the integrals, we truncate the integration limits at 5$\sigma$ from the peak of the Gaussian switching function. Since the Gaussian switching decays rapidly, this yields results consistent with the Gaussian switching having an infinite tail used in analytical expressions.}. It can be observed from the first plot of Fig.~\ref{fig:4} that the derivative of transition probability for a static detector, coupled with two point functions $W_{\psi,s} ^{'\text{HHI}}$ defined in \eqref{eq:13} and $W_{\psi,s} ^{'\text{Unruh}}$ defined in \eqref{eq:17}, becomes increasingly negative as one approaches the event horizon.  Therefore, the weak anti-Hawking effect for a static detector becomes more prominent as one approaches the event horizon \cite{antihawking, antiunruh}. However, for a freely falling detector (with the two-point functions 
$ W_{\psi,f} ^{'\text{HHI}}$ and $ W_{\psi,f} ^{'\text{Unruh}} $ defined in \eqref{eq:15} and \eqref{eq:19}), the top right plot shows that the anti-Hawking is dominant far from the horizon for HHI and Unruh states. In contrast, the bottom left plot of Fig.~\ref{fig:4} shows that a static detector with interaction Hamiltonian corresponding to conformal symmetry preserving two-point function  $W_{\psi,c} ^{\text{'HHI}}$= $W_{\psi} ^{\text{HHI}}$ defined in\eqref{confwighh} and $W_{\psi,c} ^{\text{'Unruh}}$ = 
$W_{\psi} ^{\text{Unruh}}$ defined in \eqref{confwighu}, for the range of parameter space considered throughout the paper, does not encounter any anti-Hawking effect. This property remains valid for a freely falling detector in HHI and Unruh states with  $W_{\psi,c} ^{\text{'HHI}}$= $W_{\psi} ^{\text{HHI}}$ and  $W_{\psi,c} ^{\text{'Unruh}}$ = $W_{\psi} ^{\text{Unruh}}$ respectively.

\section{Entanglement harvesting using UDW} \label{section4}
We discussed above the transition rate and the anti-Hawking effects experienced by the UDW detector. In this section, we introduce the entanglement measure, put two such detectors along a given trajectory, and perform the entanglement harvesting protocol.
\subsection{Entanglement measures}
For a state $ \rho_{AB} $ in the space of states in the Hilbert space of the combined system, an entanglement measure, denoted as E($ \rho_{AB} $ ), is defined as a mapping from the space of states to the set of non-negative real numbers ($ \mathds{R} ^{+}$). The entanglement measure must satisfy the following conditions: E($\rho_{AB}$ ) = 0 if $\rho_{AB}$ represents a separable state. Additionally, E($\rho_{AB}$) should not, on average, increase under local operations and classical communications (LOCC). Measuring entanglement is a broad and dynamic area of research in its own right, with many different approaches suggested for quantifying it. The entanglement of formation, intended to assess the resources needed for creating a particular entangled state, is recognized as one of the most fundamental measures \cite{Plenio:2007zz}. Since the entanglement of formation rises monotonically with the concurrence, determining the concurrence C of the detectors' final state suffices to measure the amount of entanglement between them \cite{Plenio:2007zz}. It provides a bound on the entanglement. The concurrence is defined by
\begin{equation} \label{eq:38}
    \text{C} [\rho_{AB} ] := \operatorname{max} [ 0, \lambda _1 - \lambda _2 - \lambda _3 - \lambda _4 ],
\end{equation}
where the $\lambda_s$ are the square roots of the eigenvalues of $\rho_{AB} \tilde{\rho}_{AB}$ in decreasing order. Here, $\tilde{\rho}_{AB}$ is defined as $(\sigma_y \otimes \sigma_y) \rho^*_{AB} (\sigma_y \otimes \sigma_y)$, and $\sigma_y$ represents the usual Pauli matrix. Substituting the $\lambda_s$ for the reduced density matrix $\rho_{AB}$, shown in \eqref{eq:28}, in Eq.\eqref{eq:38} one obtains \cite{tjoaman, gytm}
\begin{equation} \label{eq:39}
    \text{C}[\rho _{AB}] = 2 \operatorname{max} [0, |\mathcal{M}| - \sqrt{\mathcal{L}_{AA} \mathcal{L}_{BB} } ] + \mathcal{O} (\lambda ^4).
\end{equation}

Other forms of quantum correlations, such as quantum discord, which quantifies the overall quantumness of correlations, also play an important role in relativistic quantum information. We also compute the quantum mutual information, which measures the total correlation between two probes. For a quantum bipartite system, it is defined to be the relative entropy between $\rho _{AB}$ and $\rho_A \otimes \rho_B$  \cite{tjoaman, gytm, PhysRevD.92.064042}
\begin{align} \label{eq:40}
    I[\rho_{AB}] & := S(\rho _{AB}|\rho_A \otimes \rho_B ) = S(\rho_A ) + S(\rho_B ) - S(\rho _{AB}) .   
\end{align}
Here S(..) denotes the von Neumann entropy, and it is given by the following expressions: $S(\rho_{AB}) = -\operatorname{Tr}_{AB}(\rho_{AB}\log{\rho_{AB}})$, $S(\rho_{A}) = -\operatorname{Tr}_{A}(\rho_{A}\log{\rho_{A}})$, and $S(\rho_{B}) = -\operatorname{Tr}_{B}(\rho_{B}\log{\rho_{B}})$.
By expanding Eq. (\ref{eq:40}) for mutual information to the leading order in terms of the coupling strength $\lambda$, one gets
\begin{equation} \label{eq:41}
    I[\rho_{AB}] = \mathcal{L} _+ \log{\mathcal{L} _+} + \mathcal{L} _- \log{\mathcal{L} _-} - \mathcal{L} _{AA} \log{\mathcal{L} _{AA}} -\mathcal{L} _{BB} \log{\mathcal{L} _{BB}} + \mathcal{O} (\lambda ^4)
\end{equation}
where,
\begin{equation} \label{eq:42}
    \mathcal{L}_\pm = \frac{1}{2} ( \mathcal{L}_{AA} + \mathcal{L}_{BB} \pm \sqrt{ ( \mathcal{L}_{AA} - \mathcal{L}_{BB})^2 + 4 \mathcal{L} _{AB} \mathcal{L} _{BA}}) .
\end{equation}
In the next subsection, we compute the mutual information, and the concurrence for two UDW detectors kept along a given trajectory.
\subsection{Vacuum correlations outside the horizon } \label{section 42}
In the context of a free scalar field in Minkowski spacetime, when we choose to decompose the state space into plane-wave modes, the overall Hilbert space takes the form of a direct product of infinitely countable harmonic oscillator state spaces, each corresponding to a distinct mode k. Consequently, the resultant state appears as a product state rather than an entangled one. However, an alternative approach involves utilizing a tensor product of two-mode squeezed (TMS) states in pairs of Rindler modes, which gives rise to entangled Rindler wedges. Additionally, one can also represent the quantum field's state as a path-ordered or time-ordered exponential operator, acting on local degrees of freedom at each point in space, using variational ansätze like continuous matrix-product states (cMPS) and the continuous multiscale entanglement renormalization ansatz (cMERA) \cite{Ciraac,Haegeman}. Such an approach allows for the interpretation of the vacuum state as a multipartite entangled state \cite{martin3}. There could be multiple possible choices based on the natural choice for the Hilbert space decomposition, which depends upon the reference frame. Hence, the entanglement harvested by the detector is also expected to depend upon trajectory. In this section and the subsequent ones, we delve into the entanglement characteristics of the multipartite entangled state, employing the assistance of two Unruh-DeWitt (UDWs) detectors positioned at distinct locations and following various trajectories.

\begin{figure*}[ht!] 
        \centering
        \includegraphics[width=0.92\textwidth]{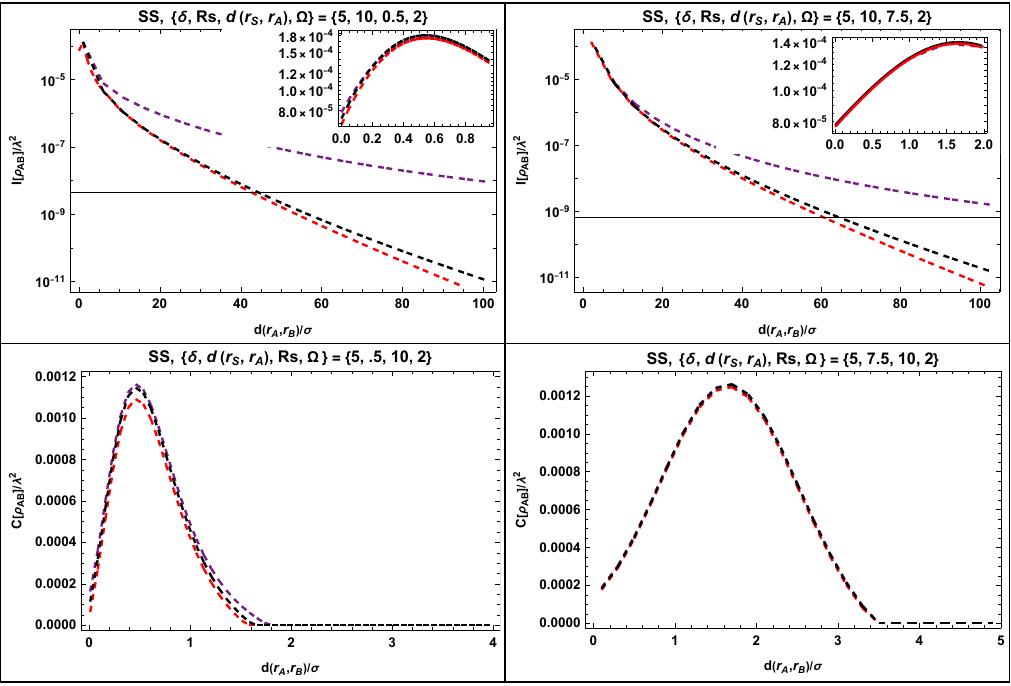}
        \captionsetup{margin=1cm, font=small}
        \caption{ The above figures show the numerical plots of the mutual information (the top panels) and the concurrence (the bottom panels) with two static UDWs as a function of the proper distance between two UDWs on a constant Painleve-Gullstrand time slice (x-axis) while keeping the distance of detector A from the event horizon and all other parameters fixed. In the plots, the black line represents the Unruh state, the red line represents the Hartle-Hawking-Israel (HHI) state, and the purple line represents the Boulware state. Here, the two-point function corresponding to the interaction Hamiltonian shown in Eq.\eqref{eq:24}, $W_{\psi,c} ^{'\alpha}$ (= $W_{\psi} ^{\alpha}$), is taken.}
        \label{fig:5}
\end{figure*} 

\subsubsection{Two static detectors at various separations} \label{subsec4.2.1}
To initiate our investigation, we position two static detectors, labeled A and B, with detector A kept at a radial coordinate of $r_A$ and detector B kept at $r_B$. Due to their differing radial coordinates, these detectors experience distinct redshift effects. We switch on detector A with the peak of the Gaussian switching function centered at a constant time slice $t_{PG}$, while the other detector B is activated with its corresponding Gaussian switching function peak delayed by a parameter $\delta$ from the constant time slice $t_{PG}$. The corresponding proper times in terms of $t_{PG}$ and $\delta$ are given by
\begin{equation} \label{eq:43}
    \tau _{A0} = \sqrt{1-\frac{R_s}{r_A}} \bigg( t_{PG} - 2R_s \sqrt{\frac{r_A}{R_s} } - R_s\log{ \bigg|\frac{\sqrt{\frac{r_A}{R_s} }-1 }{\sqrt{\frac{r_A}{R_s} }+1} \bigg| } \bigg),
\end{equation}

\begin{equation} \label{eq:44}
    \tau _{B0} = \delta + \sqrt{1-\frac{R_s}{r_B}} \bigg( t_{PG} - 2R_s \sqrt{\frac{r_B}{R_s} } - R_s \log{\bigg| \frac{\sqrt{\frac{r_B}{R_s} }-1 }{\sqrt{\frac{r_B}{R_s} }+1} \bigg| } \bigg).
\end{equation}
The tail of Gaussian switching makes it difficult to arrange both detectors to be purely spacelike separated. However, one can define a communication estimator \cite{gytm, PhysRevD.108.045015} 
\begin{equation} \label{eq:45}
    \mathcal{E} := \frac{ \lambda ^2}{2} \operatorname{Im} \bigg(  \int _{-\infty} ^{+\infty} d \tau _A  \int _{-\infty} ^{+\infty} d \tau _B \chi(\tau _A) \chi(\tau _B) \langle 0_\alpha| [:\bar{\psi} (x) \psi (x):, :\bar{\psi} (x') \psi (x') :] |0_\alpha \rangle
 \bigg),
\end{equation}
whose magnitude broadly characterizes how timelike/spacelike our detectors A and B are. The entanglement harvested by detectors gets contributions both due to the communication as well as due to the intrinsic entanglement present in the field \cite{gytm, PhysRevD.104.125005}. We choose the constant parameter $\delta$ such that the communication estimator, defined in Eq.\eqref{eq:45}, is minimized. However, taking a very large $\delta$ also reduces the magnitude of correlation measures. Hence, we need to optimize a minimum value of $\delta$ for which the $\mathcal{E}$ is close to zero. This makes the contribution received due to the communication minimal. We calculate the mutual information and concurrence for this configuration and then repeat the process for various values of the separation between detectors $d(r_A, r_B)$ while keeping $R_s$ and other parameters constant.

A static observer at infinity does not experience any flux of radiation in the Boulware vacua. However, at a finite distance from the horizon, due to the curvature of spacetime, the vacuum polarization contributes to the stress-energy tensor \cite{PhysRevD.54.5116, visser3}. In contrast to the scenario with an infinite switching duration, $\Delta \tau$, or large energy gap $\Omega$ where the excitation rate is zero, a static detector in the Boulware vacuum during a finite proper time $\Delta \tau$ experiences transient excitations resulting from the switching process  \cite{Ng:2021enc}. Therefore, one can also get a nonzero transition rate in the Boulware vacuum. By substituting the pullbacks of the two-point functions \( W_{\psi} ^{\text{HHI}} \) shown in \eqref{confwighh}, \( W_{\psi} ^{\text{Unruh}} \) shown in \eqref{confwighu}, and \( W_{\psi,s} ^{'\text{Boulware}} \) (= \( W_{\psi} ^{\text{Boulware}} \)) shown in \eqref{confwighb} along the trajectory of a static detector in Eqs.~\eqref{eq:29}, \eqref{eq:30} for $\mathcal{L}_{ij}$ and $\mathcal{M}$, and further using \eqref{eq:39}, \eqref{eq:41} for concurrence and mutual information, we obtain the numerical plots of the correlation measures displayed in Fig.~\ref{fig:5}, where we have considered the case when one detector is positioned in close proximity to the horizon, and we undertake entanglement harvesting with the second detector held static at various proper separations. Analyzing the results in plots shown in Fig.~\ref{fig:5}, it is observed that the mutual information, as well as the concurrence in all three vacua, follow the same trend for smaller separation between detectors. One can understand it as all three vacua have similar entanglement properties at small scales. However, for larger separation between detectors, the mutual information as well as the concurrence becomes maximum at a substantial proper separation in the Boulware vacuum state, while it is minimum in the Hartle-Hawking-Israel (HHI) state. The absence of Hawking radiation for a static detector in the Boulware state, coupled with the fact that we are comparing the vacuum correlations of different states at the same proper distance, where the gravitational redshift, as well as other parameters, remain identical, implies that at sufficiently large distances between detectors, the presence of Hawking radiation in the Unruh and HHI states diminish the total vacuum correlation as well as the entanglement as observed from the vantage point of static detectors \footnote{Here, one should recall that the Unruh state has only outgoing flux of Hawking radiation, while HHI state has both outgoing as well as ingoing radiation. This explains why the Unruh state correlations are in the middle.}. As we decrease the proper distance between the UDW detectors, A and B, while keeping detector A fixed near the horizon, the mutual information in all three states increases in such a way that, at smaller separations, all states follow the same trend. Additionally, if detector A is placed at a different radial coordinate, the difference in mutual information between the states—for a given separation between the UDW detectors—changes accordingly. These observations suggest that in the vicinity of the black hole horizon, the energy flux and energy density of Hawking radiation have a detrimental effect on vacuum correlations, causing their degradation.

The observations made in the preceding paragraph, based on UDW formalism, are also consistent with the resolvent technique prediction in section \ref{sec:2.4} (see the bottom panel of Fig.~\ref{fig:1}). In particular, the relative ordering of mutual information in all three states matches. We noted in section \ref{sec:2.4} (and Fig.~\ref{fig:1}) that as we decrease the separation between disjoint intervals, there exist specific points where the correlation measure reaches its maximum and then decays again. This feature can also be seen in the entanglement harvesting with UDW in Fig.~\ref{fig:5}.  We observe a peak in the mutual information plots displayed in the bottom panel of Fig.~\ref{fig:1}. This can be understood as the two disjoint intervals reaching a minimum separation. The standard width of the correlation plots in Fig.~\ref{fig:5} increases, indicating a sharper decay near the horizon. This behavior is consistent across all three states, suggesting that these features are primarily due to vacuum polarization and the gravitational redshift effect, as there is no Hawking radiation for a static observer in the Boulware state.

The results in the present subsection indicate that the presence of a horizon has a substantial impact on the vacuum correlations of a quantum field. Nonetheless, we expect other factors, such as the separation distance between detectors, the energy gap, relative velocity, the proper distance from the horizon, and spacetime curvature, to be relevant as well. To examine the dependence of this effect on the distance from the horizon, the following two subsections focus on positioning two Unruh-DeWitt detectors at a fixed separation distance at the time of the peak of switching, denoted as $d(r_A,r_B)$, and explore the entanglement dynamics across various distances from the black hole horizon along various trajectories.

\subsubsection{Keeping the difference of radial coordinate fixed ---both detectors are static (SS) } \label{subsec4.2.2}

\begin{figure*}[ht!] 
        \centering
        \includegraphics[width=0.92\textwidth]{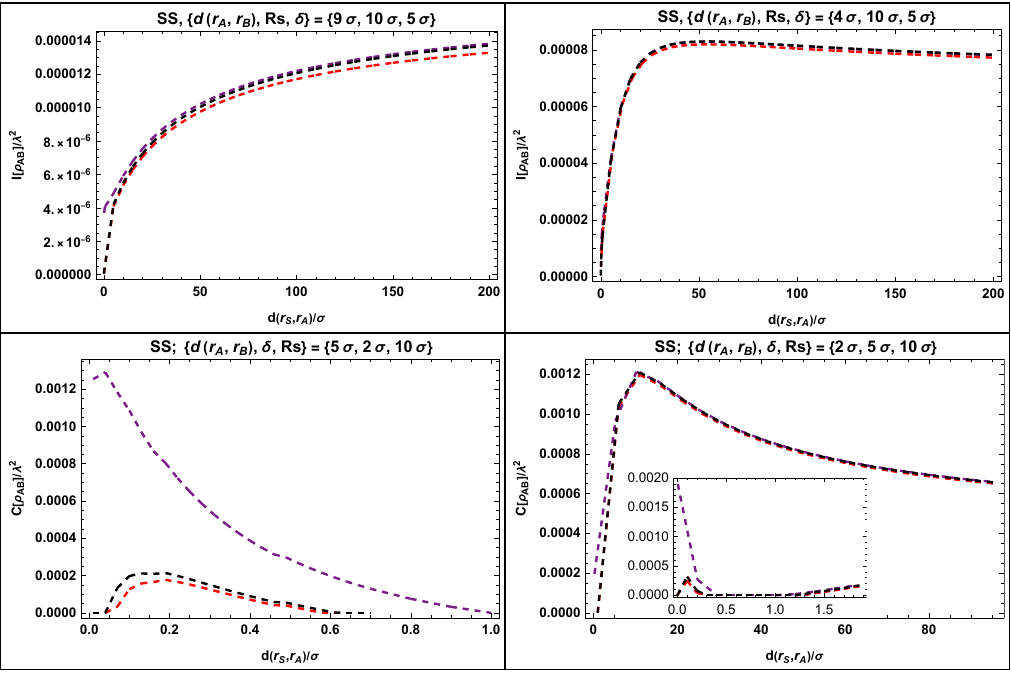}
        \captionsetup{margin=1cm, font=small}
        \caption{  The above figures show the numerical plots of the mutual information (the top panel) and the concurrence (the bottom panel) with two static UDWs as a function of the proper distance of the detector A from the event horizon on a constant Painleve-Gullstrand time slice (x-axis) while keeping the distance between two detectors and all other parameters fixed. In the plots, the black line represents the Unruh state, the red line represents the Hartle-Hawking-Israel (HHI) state, and the purple line represents the Boulware state. Here, the two-point function corresponding to the interaction Hamiltonian shown in Eq.\eqref{eq:24}, $W_{\psi,c} ^{'\alpha}$ (= $W_{\psi} ^{\alpha}$), is taken. We keep $\Omega = 2 \sigma$ in all cases. }    
        \label{fig:6}
\end{figure*}

\begin{figure*}[ht!]  
        \centering
        \includegraphics[width=0.92\textwidth]{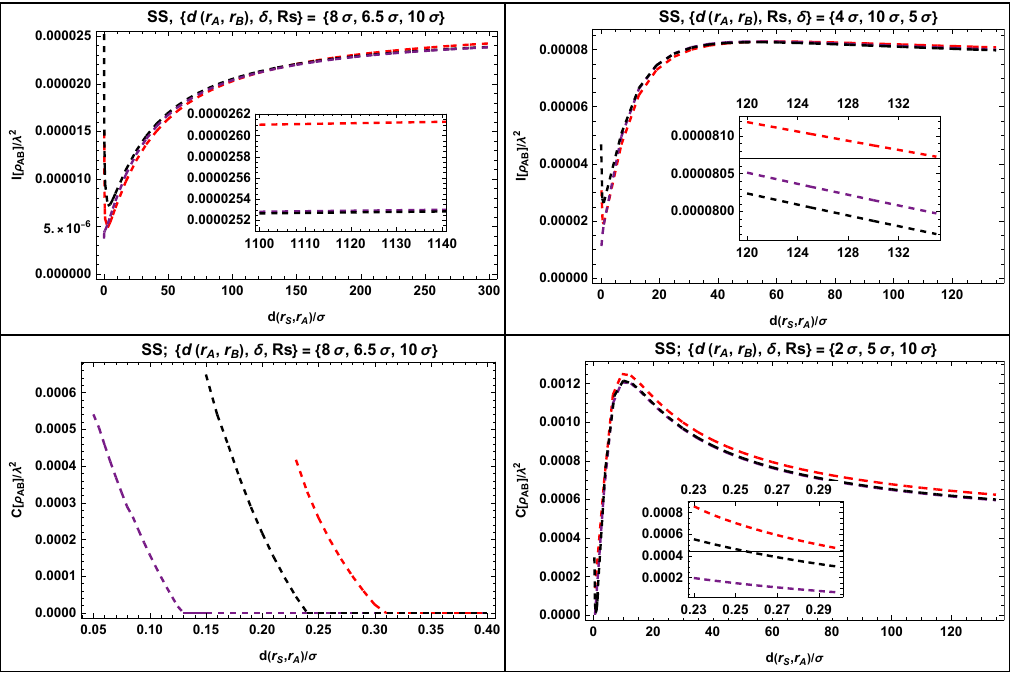}
        \captionsetup{margin=1cm, font=small}
        \caption{ The above figures show the numerical plots of the mutual information (the top panel) and the concurrence (the bottom panel) with two static UDWs, coupled to the field with interation Hamiltonian $\hat{H} _{\text{j}}^{'\text{int}}$, shown in Eq.\eqref{eq:24old}, with static 2-bein, as a function of the proper distance of the detector A from the event horizon on a constant Painleve-Gullstrand time slice (x-axis) while keeping the distance between two detectors and all other parameters fixed. In the plots, the black line represents the Unruh state, the red line represents the Hartle-Hawking-Israel (HHI) state, and the purple line represents the Boulware state. We keep $\Omega = 2 \sigma$ in all cases.  }    
        \label{fig:7}
\end{figure*}

In this subsection, we maintain all configurations identical to those in the previous subsection, with the exception that we now vary the distance of detector A from the horizon while keeping the difference in radial coordinates between both detectors, denoted as $d(r_A,r_B)$, fixed at a fixed Painlevé-Gullstrand (PG) time slice. This would correspond to the fixed proper distance between detectors in PG coordinates. The corresponding correlation measures with conformally coupled two-point functions \( W_{\psi,c}^{'\text{HHI}} \) (= \( W_{\psi}^{\text{HHI}} \)) defined in~\eqref{confwighh}, \( W_{\psi,c}^{\text{'Unruh}} \) (=\( W_{\psi}^{\text{Unruh}} \)) defined in~\eqref{confwighu}, and \( W_{\psi,c}^{'\text{Boulware}} \) (=\( W_{\psi}^{\text{Boulware}} \)) defined in~\eqref{confwighb}, as plotted in Fig.~\ref{fig:6}, exhibit the same relative ordering of correlation measures as in the preceding subsection, as well as in Fig.~\ref{fig:1}; i.e., the mutual information and the concurrence are maximal for the Boulware state and minimal for the HHI state. This again suggests that the presence of Hawking radiation in HHI and Unruh states diminishes the vacuum correlations. Furthermore, near the event horizon, the correlation measures are close to zero for the HHI and Unruh states, suggesting the gravitational redshift also leads to diminishing the vacuum correlations. In Fig.~\ref{fig:7}, we have plotted the graphs for concurrence and mutual information for the two-point functions \( W_{\psi,s} ^{'\text{HHI}} \) defined in  Eq. \eqref{eq:13}, \( W_{\psi,s} ^{'\text{Unruh}} \) defined in  \eqref{eq:17}, and \( W_{\psi,s} ^{'\text{Boulware}} \) defined in  \eqref{eq:21}. It can be observed that at any separation scale, the concurrence in all three states demonstrates a similar pattern, which remains nonzero and finite close to the horizon. Moreover, the mutual information is also nonzero near the event horizon for the Unruh and the HHI states. This behavior differs from that observed in the case of scalar fields, as analyzed in \cite{gytm}, where all forms of correlations vanish near the horizon for static detectors in both the Unruh and HHI states. The discussion in the current paragraph suggests that the vanishing of fermionic field entanglement near the horizon depends on the choice of the interaction Hamiltonian. One can choose an interaction Hamiltonian such that the anti-Hawking effect takes place, enhancing entanglement, as opposed to the Hawking effect, which degrades it. One can refer \cite{Montero}, for the state dependence of the decay of the entanglement in the Fermionic field under the infinite acceleration limit, which is the case near the horizon.

It can be seen from the bottom-right and top-right plots of Figs.~\ref{fig:6} and \ref{fig:7} that, apart from the decreasing behavior of correlation measures near the horizon, the total correlation, as well as the concurrence, exhibits a local maximum away from the horizon. This behavior is not evident from the top left plot because the mutual information for large-scale separated detectors varies slowly in comparison to the small-scale separated detectors. This contrasts with the bottom left plot, which shows that the concurrence for large-scale separated detectors decays rapidly and goes to zero as one moves away from the horizon. Near the horizon, both mutual information and concurrence for the Unruh vacuum closely resemble those of the HHI state, while farther from the horizon, these measures align more closely with the Boulware state. The plots in Fig.~\ref{fig:7} also suggest that, far from the horizon, the concurrence is minimized for the Boulware state and maximized for the HHI state. However, Fig.~\ref{fig:6} show that mutual information and concurrence are minimized for the HHI state and maximized for the Boulware state. This reversal suggests that the anti-Hawking effect, observed for the two-point function used in Fig.~\ref{fig:7} (see Fig.~\ref{fig:4}), enhances entanglement. The HHI state's correlation measures for the case considered in Fig. \ref{fig:7} are maximized due to the dominance of the anti-Hawking effect far from the horizon, whereas for the case of the two point function corresponding to conformal symmetry preserving frame tetrad in Fig.~\ref{fig:6}, Hawking radiation diminishes correlations in the HHI state.

One also observes that for smaller separation between detectors, all plots of Fig~\ref{fig:6} follow a similar pattern, which is consistent with the observation in the previous subsection \ref{subsec4.2.1} that the entanglement properties on a small scale follow a similar trend in all vacua. Additionally, as the detectors are positioned at greater separations, the concurrence diminishes, especially at larger distances from the horizon. Here, by diminishes, we mean that the concurrence becomes small enough to implement a harvesting protocol with this energy gap. The observations of this section imply that the anti-Hawking effect, in conjunction with Hawking radiation and the gravitational redshift, plays a pivotal role in determining the entanglement properties in the vicinity of a black hole. To explore whether the decay of all correlations in the plots near the horizon is due to the high gravitational redshift and to investigate the role of the anti-Hawking effect further, we consider two freely falling detectors in the next subsection.

\subsubsection{Two freely falling detectors (FF)} \label{subsec4.2.3}
In the realm of quantum physics, a unique characteristic emerges, where the structure of the Hilbert space can undergo substantial variations among different observers, and the nature of entanglement is contingent upon how one decomposes the state space. The process of black hole evaporation is a physical phenomenon, and, as such, the mass of the black hole must diminish from the perspective of any observer. Consequently, a freely falling timelike detector in the HHI and Unruh states will also perceive the presence of radiation (which need not be thermal) from the black hole \cite{Shankaranarayanan:2000gb, gytm, Barbado:2011dx, Dey:2019ugf}. Additionally, a freely falling detector will also perceive a time-dependent vacuum polarization in all three vacua. These facts also motivate us to envisage both detectors as being in free fall from infinity, with Gaussian switching function peaked at different times. We choose detectors to be initially at rest at spatial infinity so that the adapted coordinate system is the Gullstrand–Painlevé coordinates. In this scenario, both the proper distance between the detectors and the gravitational redshift are subject to change over time. Additionally, the local temperature perceived by detectors, defined in Eq.\eqref{teffdef}, also changes due to the combination of Hawking and anti-Hawking effects. 

\begin{figure*}[ht!] 
        \centering
         \includegraphics[width=0.92\textwidth]{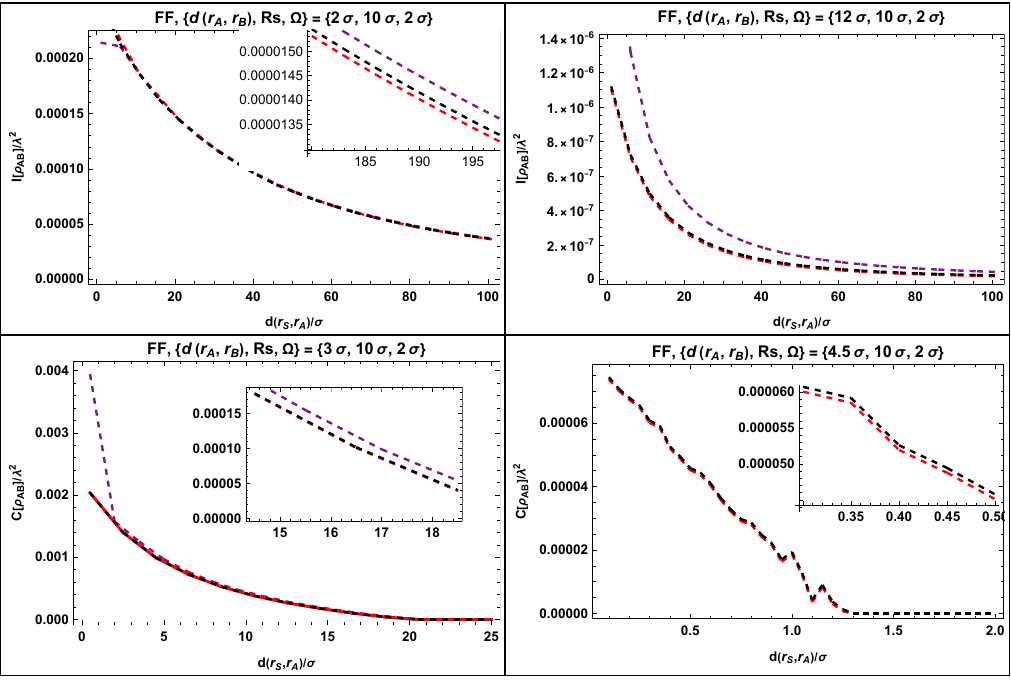}
         \captionsetup{margin=1cm, font=small}
        \caption{ The above figures show the numerical plots of the mutual information (the top panel) and the concurrence (the bottom panel) with two freely falling UDWs as a function of the proper distance of detector A from the event horizon on a constant Painleve-Gullstrand time slice (x-axis) while keeping the distance between two detectors at the peak of Gaussian switching and all other parameters fixed. In the plots, the black line represents the Unruh state, the red line represents the Hartle-Hawking-Israel (HHI) state, and the purple line represents the Boulware state.  Here, the two-point function corresponding to the interaction Hamiltonian shown in Eq.\eqref{eq:24}, $W_{\psi,c} ^{'\alpha}$ (= $W_{\psi} ^{\alpha}$), is taken. We keep $\Omega = 2 \sigma$ in all cases.  }  
        \label{fig:8}
\end{figure*}

\begin{figure*}[ht!] 
         \centering
         \includegraphics[width=0.92\textwidth]{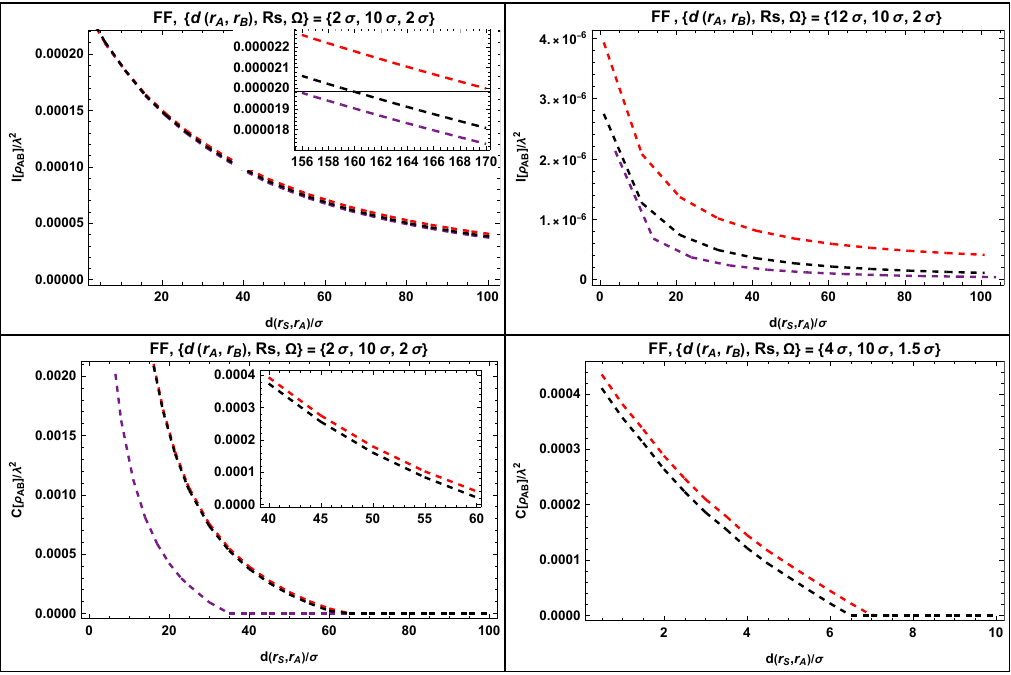}
         \captionsetup{margin=1cm, font=small}
        \caption{ The above figures show the numerical plots of the mutual information (the top panel) and the concurrence (the bottom panel) with two freely falling UDWs as a function of the proper distance of detector A from the event horizon on a constant Painleve-Gullstrand time slice (x-axis) while keeping the distance between two detectors at the peak of Gaussian switching and all other parameters fixed. In the plots, the black line represents the Unruh state, the red line represents the Hartle-Hawking-Israel (HHI) state, and the purple line represents the Boulware state. The interaction Hamiltonian \( \hat{H} _{\text{j}}^{'\text{int}} \) shown in Eq.~\eqref{eq:24old} is used with the freely falling 2-bein, and we keep \( \Omega = 2\sigma \) in all cases. }  
        \label{fig:9}
\end{figure*}

We consider an ensemble of pairs of radially freely falling detectors and perform entanglement harvesting when the separation between the two detectors approaches a proper separation \( d(r_A, r_B) = r_B - r_A \) (at constant Painleve Gulstrand time slice), where, $r_A$ and $r_B$ are related to the time of peak \( \tau_{A0} \) and \( \tau_{B0} \) of the respective Gaussian switching function, as defined in Eq.~\eqref{eq:46}. The Gaussian switching is centered at the proper times of the individual detectors, given by
\begin{eqnarray} \label{eq:46} 
    \tau _{A0} = - \frac{r_A}{3} \sqrt{\frac{4 r_A}{R_s}}
&&   ;  \tau _{B0} = - \frac{r_B}{3} \sqrt{\frac{4 r_B}{R_s}} 
\end{eqnarray}
\\
which is precisely the time taken by a freely falling test particle to arrive at $r_A$ or $r_B$, respectively, starting from infinity with zero initial velocity. The entanglement harvesting is repeated with this setup multiple times, varying the distance of the switching function peak of the nearest detector from the horizon, while keeping \( d(r_A, r_B) \) same for all pairs. The results are shown in Figs.~\ref{fig:8} and \ref{fig:9}.  

It's worth noting that in the case of the Boulware vacuum, we refrain from conducting measurements very close to the horizon, as the freely falling detector must inevitably cross the horizon, and the state is not well-defined in that region. Hence, we can compare entanglement measures at locations far away from the horizon for the Boulware vacuum.

We have employed a pointlike detector, which means that in the case of free fall, it does not locally experience any gravitational field. This characteristic allows us to collect entanglement data very near the black hole horizon. Furthermore, in the reference frame of the freely falling detector, the effective temperature, defined in Eq.\eqref{teffdef}, at the horizon remains finite \cite{Barbado:2011dx}, as the observer in this frame is not subjected to acceleration locally. Therefore, we do not expect the correlations to decay near the horizon when observed from the perspective of freely falling detectors in any of these vacuum states.

\newpage
From Figs.~\ref{fig:2}, \ref{fig:8}, and \ref{fig:9}, it is evident that the qualitative behavior of mutual information and concurrence in each of these three vacuum states are quite similar. In particular, we do not observe any decay in correlation measures near the horizon. This observation suggests that the decay near the horizon, observed in subsection \ref{subsec4.2.2}, is due to high gravitational acceleration.

In Fig.~\ref{fig:9}, we have plotted the mutual information and concurrence with two-point functions $ W_{\psi,f} ^{'\text{HHI}}$ defined in \eqref{eq:15}, $ W_{\psi,f} ^{'\text{Unruh}}$ defined in \eqref{eq:19}, and $ W_{\psi,f} ^{'\text{Boulware}}$ defined in \eqref{eq:23}. One observes in Fig.~\ref{fig:9} that both the mutual information and the concurrence are maximum for the HHI vacuum, while they attain their minimum values for the Boulware vacuum. This ordering can be understood by the enhancement of correlations due to the dominance of the anti-Hawking effect in the HHI state far from the horizon (see Fig~\ref{fig:4}). This ordering differs to that observed in Fig.~\ref{fig:2} and  Fig.\ref{fig:8} with conformal symmetry obeying two-point functions $ W_{\psi,c} ^{'\text{HHI}}$ (=$ W_{\psi} ^{\text{HHI}}$) defined in  \eqref{confwighh}, $ W_{\psi,c} ^{'\text{Unruh}}$ (=$ W_{\psi} ^{\text{Unruh}}$) defined in  \eqref{confwighu} and $ W_{\psi,c} ^{\text{'Boulware}}$ (=$ W_{\psi} ^{\text{Boulware}}$) defined in  \eqref{confwighb}. The ordering pattern is the same for both small and large-scale initial separations. Such a behavior can be attributed to the fact that the freely falling detectors need to traverse through the field ``bath" (need not be a thermal bath), and in the case of the Unruh vacuum, there is only an outgoing flux of radiation \cite{gytm, Barbado:2011dx}. Meanwhile, in the HHI vacuum, there are both ingoing and outgoing fluxes of radiation (need not be thermal). This difference in radiation fluxes immediately leads to more entanglement disruption in the HHI vacuum.  It's important to highlight that, while keeping the peak of detector A very close to the horizon, the detector does spend some time beyond the horizon due to the long tail of the Gaussian switching function. However, as evident from the plots for both the Unruh and HHI vacua, they exhibit smooth behavior near the horizon. This suggests that the detectors do not experience any sudden or abrupt changes as they cross the black hole horizon. We also note that both the mutual information and the concurrence are finite and smooth as one approaches the event horizon in all cases. We next explore the near-horizon entanglement properties in the next section.
\section{Near horizon entanglement of HHI vacuum } \label{section5}

\begin{figure*}[ht!] 
         \centering
         \includegraphics[width=.92\textwidth]{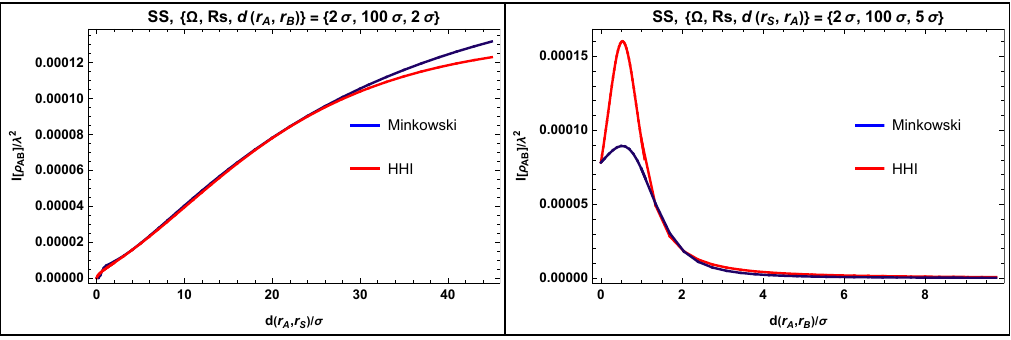}
         \captionsetup{margin=1cm, font=small}
        \caption{Plots for the comparison of Minkowski and HHI states mutual information near the horizon with two static detectors having coupling $\hat{H} _{\text{j}}^{\text{int}}$ shown in Eq.\eqref{eq:24}. }  
        \label{fig:10}
\end{figure*}

The proper distance between two Schwarzschild coordinates at a fixed Schwarzschild time slice and constant $\theta$ and $\phi$ is given by 
\begin{equation} \label{eq:47}
    z = \int _{r_A} ^{r_B} \frac{1}{\sqrt{1-\frac{R_s}{r}}} dr .
\end{equation}
In terms of the proper distance z defined above, the truncated Schwarzschild metric near the event horizon is given by
\begin{equation} \label{eq:48}
    ds ^2 \approx - z ^2 dt_S ^2 + dz ^2 .
\end{equation}
In terms of $\kappa \equiv (1/2)f'(r=R_s)$ and  $l \equiv$ ($r$ - $R_s$), with $ f(r) = (1-R_s/r), $  the above metric in Eq. \eqref{eq:48} is expressed as
\begin{equation} \label{eqn:48}
    ds ^2 \approx - 2 \kappa l dt_S ^2 + \frac{dl ^2}{2\kappa l} .
\end{equation}
This is precisely the metric of a Rindler observer in flat Minkowski spacetime. The stress-energy tensor for the Unruh vacua exhibits divergences near the past horizons. While the stress-energy tensor for the Boulware vacuum has divergence on both future and past event horizons. The divergence underscores the significance of considering the influence of the matter sector in shaping the spacetime geometry, signaling a deviation from the semiclassical regime in the vicinity of the horizon. In contrast, there is no such divergence for the HHI state. In a flat spacetime scenario, the expectation value of stress energy in the Rindler vacuum is divergent near the horizon, whereas no such divergence is present for the expectation value in the Minkowski vacuum. The equivalence principle implies that the local gravitational effect felt by a static detector is equivalent to an acceleration in Minkowski spacetime. Furthermore, a uniformly accelerating detector in flat space experiences Unruh radiation, analogous to the Hawking radiation encountered by a static detector in the HHI state. These observations prompt us to draw a comparison between the near-horizon properties of the HHI vacuum for a static observer and the Minkowski vacuum for a uniformly accelerated observer, i.e., a Rindler observer. 

From Eq.\eqref{mutualmain}, the mutual information between two disjoint intervals in the HHI vacuum is given by
\begin{eqnarray} \label{eqn:64}
I_{\text{HHI}} = \frac{1}{12} \log{ \left( \frac{ (V_1 - V_3)^2(U_1 - U_3)^2 (V_2 - V_4)^2(U_2 - U_4)^2}{ (V_3 - V_2)^2(U_3 - U_2)^2 (V_1 - V_4)^2(U_1 - U_4)^2} \right) } .
\end{eqnarray}
In the limit, when the interval A is very close to the future event horizon $U_1 \rightarrow 0$ and $U_2 \rightarrow 0$. Therefore, Eq. \eqref{eqn:64} becomes
\begin{eqnarray} \label{eqn:65}
I_{\text{HHI}} \approx \frac{1}{12} \log{ \left( \frac{ (V_1 - V_3)^2 (V_2 - V_4)^2}{ (V_3 - V_2)^2 (V_1 - V_4)^2} \right) } .
\end{eqnarray}
In the same limit $V = 2 R_s e^{v/2R_s}$ $\rightarrow$ $2 R_s^{1/2} e^{t_S/2r_s}\sqrt{r-R_s}$. Considering intervals on a $t_S$=constant hypersurface, Eq.~\eqref{eqn:65} can be rewritten as
\begin{eqnarray} \label{eqn:66}
I_{\text{HHI}} \approx \frac{1}{12} \log{ \left( \frac{ (\sqrt{r_1 - R_s} - \sqrt{r_3 - R_s})^2 (\sqrt{r_2 - R_s} - \sqrt{r_4 - R_s})^2}{ (\sqrt{r_3 - R_s} - \sqrt{r_2 - R_s})^2 (\sqrt{r_1 - R_s} - \sqrt{r_4 - R_s})^2} \right) } .
\end{eqnarray}
The mutual information between two disjoint intervals $[(U_{M1}, V_{M1}), (U_{M2}, V_{M2})]$ and $[(U_{M3}, V_{M3}), (U_{M4}, V_{M4})]$ for a Rindler observer in Minkowski vacuum
, with $(U_{Ms},V_{Ms})$ being the Minkowski null coordinates, is given by
\begin{eqnarray} \label{eqn:67}
I_{\text{Mink}} = \frac{1}{12} \log{ \left( \frac{ (V_{M1} - V_{M3})^2(U_{M1} - U_{M3})^2 (V_{M2} - V_{M4})^2(U_{M2} - U_{M4})^2}{ (V_{M3} - V_{M2})^2(U_{M3} - U_{M2})^2 (V_{M1} - V_{M4})^2(U_{M1} - U_{M4})^2} \right) } .
\end{eqnarray}
In the near horizon limit, $U_{M1} \rightarrow 0$ , $U_{M2} \rightarrow 0$ and $V_{Ms}$ $\rightarrow$ $\sqrt{2l_s} e^{g\tau}/\sqrt{g}$. Here `g' is the acceleration parameter of the Rindler observer. Furthermore, taking both intervals on $\tau $ = constant hypersurface Eq.\eqref{eqn:67} becomes
\begin{eqnarray} \label{eqn:66}
I_{\text{Mink}} \approx \frac{1}{12} \log{ \left( \frac{ (\sqrt{l_1} - \sqrt{l_3})^2 (\sqrt{l_2} - \sqrt{l_4})^2}{ (\sqrt{l_3} - \sqrt{l_2})^2 (\sqrt{l_1} - \sqrt{l_4})^2} \right) } .
\end{eqnarray}
Identifying  $l_s$ = $r - R_s$ we get $I_{\text{Mink}} \approx I_{\text{HHI}}$. Therefore, near the horizon, the mutual information between two nearby disjoint intervals in both situations is equivalent.

To see the equivalence of the entanglement properties from the point of view of an UDW detector, let us consider two UDW detectors, namely A and B, along the following trajectories in the Minkowski vacuum.
\begin{align} \label{eq:49}
   \bigg\{ \text{Detector A: } t_A = \frac{1}{g_A} \sinh{g_A \tau_A} ;  x_A = \frac{1}{g_A}   \cosh{g_A \tau_A}  \bigg\} \\  \bigg\{  \text{Detector B: } t_B = \frac{1}{g_B}   \sinh{g_B \tau_B} ;  x_B = \frac{1}{g_B}   \cosh{g_B \tau_B}   \bigg\} \label{eq:50}
\end{align}
\\
where $\tau_A$ and $\tau_B$ are proper time in the frame of detectors A and B, respectively, x's and t's are their Minkowski coordinates, and $g_A$ and $g_B$ are their accelerations. We take events' common line of simultaneity in both detector frames to pass from the origin of chosen coordinates, namely
\begin{equation} \label{eq:51}
    \frac{t_A}{x_A} = \frac{t_B}{x_B}
\end{equation}
Imposing the above constraints on trajectories Eqs. \eqref{eq:49} and \eqref{eq:50} we get
\begin{equation} \label{eq:52}
    \tanh{g_A \tau_A} = \tanh{g_B \tau_B}
\end{equation}
Now, since the arguments are real, we can utilize the fact that hyperbolic functions are one-to-one to write $g_A \tau_A = g_B \tau_B$. Therefore, the trajectory becomes
\begin{align} \label{eq:53}
   \bigg\{ \text{Detector A: } t_A = \sqrt{\frac{2 l_A}{g}} \sinh{g_A \tau_A} ;  x_A = \sqrt{\frac{2 l_A}{g}} \cosh{g_A \tau_A}  \bigg\} \\  \bigg\{  \text{Detector B: } t_B = \sqrt{\frac{2 l_B}{g}} \sinh{g_A \tau_A} ;  x_B = \sqrt{\frac{2 l_B}{g}} \cosh{g_A \tau_A}   \bigg\} 
\end{align} \label{eq:54}
where $l_s$ describe the Rindler coordinates and $g_s = \sqrt{g/2l_s}$. The proper distance between detectors, say, at $t_A = t_B = 0$, is given by 
\begin{equation*}
    - \frac{\sqrt{2 g l_A}}{g} + \frac{\sqrt{2 g l_B}}{g}.
\end{equation*}
We choose $\l_A$ = 1/2g and fix $\l_B$ such that the above proper distance corresponds to the same proper distance between detectors in the Schwarzschild metric. Therefore, equating this with Eq.  (\ref{eq:35}) we get,
\begin{equation} \label{eq:55}
    l_B = \frac{1}{2g}  \bigg( 1 + g \int _{r_A} ^{r_B} \frac{1}{\sqrt{1-\frac{R_s}{r}}} dr     \bigg)^2
\end{equation}
The parameter g is fixed by taking the following local acceleration felt by detector A at $r_A$ in Schwarzschild spacetime:
\begin{equation} \label{eq:56}
    g = \frac{R_s}{2 r_A ^2 \sqrt{1 - \frac{R_s}{r_A}}}.
\end{equation}
\\
With the above parameters and the two point function $ W_{\psi} ^{\text{HHI}}$ shown in Eq.\eqref{confwighh}, we repeat the entanglement harvesting protocol discussed in section \ref{section 42} for two static UDWs in HHI and two uniformly accelerating UDWs in the Minkowski vacuum. The result is displayed in Fig. \ref{fig:10}. One observes from Fig. \ref{fig:10} that the total correlation near the horizon in both of these situations, namely the HHI state for a static observer near the horizon and the Minkowski vacuum for a uniformly accelerated observer follows a similar trend. It is also apparent that as we move farther from the horizon, the disparity in correlations between the two scenarios increases. This observation suggests that the similarity in correlations between the two situations is primarily limited to the vicinity of the horizon.

\section{Discussion}
We investigated the entanglement features of a massless fermionic field outside a Schwarzschild black hole in various states, namely the Hartle-Hawking-Israel (HHI), Unruh, and Boulware states.  The entanglement observed by an observer using a detector depends on several parameters and in order to understand what an observer measures, we employed the entanglement harvesting protocol to study the correlations present in a massless fermionic field, the results of which are summarized below.

As a preliminary to entanglement harvesting, we first analyzed the transition rate of the detector. Specifically, for a static Unruh–DeWitt (UDW) detector coupled to the scalar density of a massless fermionic field—even when it is coupled to the rescaled Hamiltonian $\hat{H} _{\text{j}}^{'\text{int}}$ shown in Eq.~\eqref{eq:24old}—the transition rate in the HHI state is found to follow a Planckian distribution, as expected.  A similar analysis for the UDW detectors coupled with the scalar density of the Fermionic field in the Unruh state yields the Helmholtz free energy of a thermal bath of fermionic fields. However, the same with detectors coupled with rescaled interaction $\hat{H} _{\text{j}}^{'\text{int}}$ shown in Eq.\eqref{eq:24old} for the Unruh state yields the Helmholtz free energy of a thermal bath of fermionic or bosonic fields, depending on whether the 2-bein of the chosen interaction Hamiltonian represents 2-velocity of the local Lorentz frame preserving conformal symmetry or is selected based on the requirement that its 2-velocity match the UDW detector's velocity respectively. The presence of the anti-Hawking effect is found to be dependent on the choice of the interaction Hamiltonian through the 2-bein. If conformal symmetry is broken by the choice of the 2-bein of the interaction Hamiltonian to move with the detector, the anti-Hawking effect arises, reaching its maximum for the HHI state. In particular, for static detectors, the weak anti-Hawking effect intensifies as one approaches the horizon, while for freely falling UDW detectors, the anti-Hawking effect occurs away from the horizon and vanishes near it. In contrast, we found that detectors coupled to the scalar density of the field—whether through the interaction Hamiltonian $\hat{H} _{\text{j}}^{\text{int}}$ shown in Eq.~\eqref{eq:24}, or the rescaled interaction Hamiltonian $\hat{H} _{\text{j}}^{'\text{int}}$ with a conformal-preserving 2-bein—do not experience any anti-Hawking effect.

Using the resolvent technique, we demonstrated that for two sufficiently separated, disjoint regions, the mutual information is minimized in the HHI state and maximized in the Boulware state, provided that the separation between the intervals satisfies specific conditions stated in Appendix [\ref{Appendix E}]. The mutual information between two disjoint regions is known to be invariant under an arbitrary Weyl transformation. If we vary the proper distance between two disjoint intervals, $d(r_A,r_B)$, outside the horizon, then we get a peak in the mutual information calculated using the resolvent technique. The peak corresponds to the position where both regions are almost null-separated. One can control the distance between peak and horizon by varying \( \delta_L \), which determines the spatial extent of intervals chosen in the resolvent technique calculation. With these two disjoint intervals, when we compare two UDWs which peak around these intervals approximately and perform entanglement harvesting,  we get a single broad peak in the mutual information. This can be interpreted as, although the detector is spatially point-like, it is switched on with a Gaussian switching, and hence the temporal extent in the case of UDW and resolvent technique intervals is not the same. The width of the peaks in all three states is larger when detector A is far from the horizon compared to when it is near the horizon. This can be interpreted as the effects of vacuum polarization and gravitational redshift being large near the horizon, and as a result, the peak dies down quickly. The relative ordering of observed correlations in the UDWs for the three vacua, which matches with those in resolvent technique calculations, suggests that Hawking radiation diminishes entanglement.

Next, we fixed the distance between two detectors and varied the proper distance from the horizon of the detector closest to the horizon at a fixed Painleve-Gullstrand time slice. The relative ordering of correlation measures for different states for the case of rescaled interaction Hamiltonian $\hat{H} _{\text{j}}^{'\text{int}}$ with a 2-bein moving with the static or freely falling UDW is also found to be the reverse of that of both the conformal symmetry-preserving case with $\hat{H} _{\text{j}}^{'\text{int}}$ as well as the scalar density Hamiltonian $\hat{H} _{\text{j}}^{\text{int}}$. These observations can be interpreted as the anti-Hawking effect enhances entanglement, since the anti-Hawking effect is absent in the conformal symmetry-preserving case. While the relative ordering of the correlation measures for the case of a 2-bein moving with the static or freely falling UDW suggests that the Hawking radiation diminishes the entanglement.
More interestingly, for static detectors with static 2-bein, although some decay of entanglement measures is found near the horizon for all states, we do not get a complete death of entanglement near the horizon.  However, the decay is larger, with the conformal symmetry obeying the two-point function $ W_{\psi,c} ^{\text{State}}$, and the system of UDWs, in this case, harvests no entanglement near the horizon in HHI and Unruh states. We further confirm that the freely falling detectors don't suffer the phenomenon of entanglement decay near the horizon. In contrast, the harvested entanglement monotonically increases as one moves towards the horizon in the case of freely falling detectors. This behavior is shown to match the resolvent technique calculation.  This suggests that another cause of the entanglement decay for the static detectors near the horizon is the local gravitational acceleration felt by the detectors. 

We further showed that the near-horizon entanglement features of the HHI vacuum for a static observer, using both the resolvent technique in QFT as well as the UDW detectors numerical calculations,  are identical to those for a uniformly accelerating observer in the Minkowski vacuum in 1+1 dimensions. Entanglement features beyond the event horizon shall remain a topic for future investigation. Here, we restricted our analysis to 1+1 dimensions throughout this manuscript. However, it would be interesting to extend our study to higher dimensions and explore spacetimes beyond Schwarzschild. Notably, the presence of the anti-Hawking effect, which enhances entanglement, depends on the choice of the interaction Hamiltonian. The explicit role of the effective potential, curvature, and kinematics in the anti-Hawking effect remains an open possibility.

\section*{Acknowledgements}
The authors thank Jorma Louko for his valuable discussions and insightful comments on this manuscript.

\begin{appendices}

\section{Computing the two point function $W_\psi ^\alpha$}  \label{Appendix A}

In this appendix, we aim to compute the following normal ordered two-point function for the scalar density of the Fermionic field in the (1+1) dimensional Schwarzschild spacetime:
\begin{eqnarray} \label{eq:57}
    W_{\psi} ^{\alpha} (x,x') = \langle 0_\alpha|:\bar{\psi}_a (x) \psi _a (x) : :\bar{\psi}_b (x') \psi _b (x') : |0_\alpha \rangle ,
\end{eqnarray}
where each repeated spinor index is summed over \cite{jorma, Mandl:1985bg}. The index $\alpha$ is used to denote vacua which is being used. It is known that in two dimensions, the conformal anomaly, as well as Hawking flux, are identical for fields with spins of 0 and 1/2, as discussed in reference \cite{birrell}. Also, there is no gravitational anomaly for the Fermionic field in 1+1 spacetime dimensions\cite{parker}. Decomposing the field into positive and negative frequency parts and using normal ordering \cite{jorma, Mandl:1985bg}, we get 
\begin{align*}
   W_{\psi} ^{\alpha} (x,x') &=  \langle 0_\alpha|\bar{\psi}_a ^+ (x) \psi _a  ^+ (x) \bar{\psi}_b ^- (x')   \psi _b  ^- (x')  |0_\alpha \rangle \\
 &=  \langle 0_\alpha|\bar{\psi}_a ^+ (x) \psi _b  ^- (x') \psi _a  ^+ (x) \bar{\psi}_b  ^- (x')    |0_\alpha \rangle  \\
 &=   \langle 0_\alpha|\{ \bar{\psi}_a ^+ (x) , \psi _b  ^- (x')\} \{ \psi _a  ^+ (x), \bar{\psi}_b ^-(x')    \} |0_\alpha \rangle  \\
 & = i S_{ba} ^- (x',x) i S_{ab} ^+ (x,x') \\
  &= - Tr[S^+ (x,x') S^- (x',x) ] \\
 & = - Tr[\tilde{\gamma} ^\mu (x) U(x,x') \tilde{\gamma} ^\nu (x') U(x',x) \nabla _\mu ^{x} W_\phi ^\alpha  (x,x')  \nabla _\nu ^{x'} W_\phi ^\alpha  (x,x') ] \numberthis
 \label{templabel2}
\end{align*}
Here, we have used $\langle 0_\alpha| \psi _b ^-$ = 0 , $\psi _a ^+ |0_\alpha \rangle$ = 0 in the third line and Eq.(\ref{eq:11}) with m =0 in the in the $6^{th}$ line. Furthermore, we assume throughout that $\tilde{\gamma}^\mu \nabla_\mu U(x,x')=0$ for the parallel transport along the geodesic. Using $U(x,x') \tilde{\gamma} ^\nu (x') U(x',x) = \tilde{\gamma} ^\rho (x) \Lambda^\nu_{\ \rho}(x',x) $, where $\Lambda^\nu_{\ \rho}(x',x)$ represents the vector parallel propagator \cite{Muck2000, Allen1986, Letsios:2020twa, Avramidi2000} one gets
\begin{align*} 
 W_{\psi} ^{\alpha} (x,x') &=   - Tr[\tilde{\gamma} ^\mu (x)  \tilde{\gamma} ^\rho (x) \Lambda ^\nu _\rho (x',x)  \nabla _\mu ^{x} W_\phi ^\alpha  (x,x')  \nabla _\nu ^{x'} W_\phi ^\alpha  (x,x') ]   \\
& =  - Tr[ b_\delta ^{.\mu} (x) {\gamma} ^\delta    b_\beta ^{.\rho} (x) \gamma ^\beta \Lambda ^\nu _\rho (x',x) \nabla _\mu ^{x} W_\phi ^\alpha  (x,x')  \nabla _\nu ^{x'} W_\phi ^\alpha  (x,x') ] \\
& =  - Tr[ b_\delta ^{.\mu} (x)   b_\beta ^{.\rho} (x) \eta ^{\delta \beta } \mathds{1} \Lambda ^\nu _\rho (x',x) \nabla _\mu ^{x} W_\phi ^\alpha  (x,x')  \nabla _\nu ^{x'} W_\phi ^\alpha  (x,x') ]
\end{align*}
Here, \( b(\ldots) \) denotes the 2-bein, which has been used to express the curved spacetime Dirac gamma matrices \( \tilde{\gamma}^\mu \) in terms of the flat spacetime gamma matrices \( \gamma^\delta \) in the second line. In the third line, we have used the anticommutator relation of Dirac matrices $\{  \gamma^\delta,  \gamma^\beta \}  = 2 \mathds{1} \eta ^{\delta \beta} $. Since all indices are contracted, we can write everything else other than $\mathds{1}$ as $C$ to give the following:
\begin{align*}
    W_{\psi} ^{\alpha} (x,x') & = - Tr[C \mathds{1}] \\
& = - N b_\delta ^{.\mu} (x)   b_\beta ^{.\rho} (x) \eta ^ {\delta \beta} \Lambda ^\nu _\rho (x',x)\partial _\mu ^x W_\phi ^\alpha  (x,x')  \partial _\nu ^{x'} W_\phi ^\alpha  (x,x') . \numberthis \label{generaltwo} \\
& = - N  g ^ {\mu \rho} (x)  \Lambda ^\nu _\rho (x',x) \partial _\mu ^x W_\phi ^\alpha  (x,x')  \partial _\nu ^{x'} W_\phi ^\alpha  (x,x') \numberthis \label{generaltwonew} 
\end{align*}
Since $W_\phi^\alpha$ represents the Wightman function of a scalar field in the corresponding vacuum, we can write the spinor covariant derivative as a partial derivative, and use the identity $Tr[C \mathds{1}] = N C $, where N is the dimensionality of the spacetime, to evaluate the trace and get the second line. In the final expression, \eqref{generaltwonew}, we use $b_\delta ^{.\mu} (x) b_\beta ^{.\rho} (x) \eta ^ {\delta \beta} = g ^ {\mu \rho} (x)$.

The Wightman function of a real massless scalar field in different states are given by \cite{birrell}
\begin{align*}
    W_\phi ^{\text{Boulware}} =& - \frac{1}{4 \pi} \ln{[- \zeta ^2 (\Delta u - i \epsilon)(\Delta v - i \epsilon)]} \numberthis \label{eq:59}\\
     W_\phi^{\text{Unruh}} =& - \frac{1}{4 \pi} \ln{[- \zeta ^2 (\Delta U - i \epsilon)(\Delta v - i \epsilon)]} \numberthis \label{eq:60} \\
      W_\phi ^{\text{HHI}} =& - \frac{1}{4 \pi} \ln{[- \zeta ^2 (\Delta U - i \epsilon)(\Delta V - i \epsilon)]} \numberthis \label{eq:61}
\end{align*}
 where, $\zeta > 0$ is an IR cutoff. Since the two-point function for the scalar density of the massless spin-1/2 field is infrared-safe, we can ignore the parameter $\Lambda$ for further calculations. By repeating the above calculation for the two point function shown in Eq.\eqref{deftwopointold} one gets
 \begin{align*}
   W_{\psi,b} ^{'\alpha} (x,x') &=  \langle 0_\alpha|\bar{\psi'}_a ^{+} (x) {\psi'}_a  ^{+} (x) \bar{\psi'}_b ^- (x')   {\psi'} _b  ^- (x')  |0_\alpha \rangle \\
 &=  \langle 0_\alpha|\bar{\psi'}_a ^+ (x) {\psi'} _b  ^- (x') {\psi'} _a  ^+ (x) \bar{\psi'}_b  ^- (x')    |0_\alpha \rangle  \\
 &=   \langle 0_\alpha|\{ \bar{\psi'}_a ^+ (x) , {\psi'} _b  ^- (x')\} \{ {\psi'} _a  ^+ (x), \bar{\psi'}_b ^-(x')    \} |0_\alpha \rangle  \\
  &= - Tr[\bar{A}(x) A(x') S^+ (x,x') A(x) \bar{A}(x') S^- (x',x) ] \\
 & = - Tr[\bar{A}(x) A(x')\tilde{\gamma} ^\mu (x) U(x,x') A(x) \bar{A}(x') \tilde{\gamma} ^\nu (x') U(x',x) \nabla _\mu ^{x} W_\phi ^\alpha  (x,x')  \nabla _\nu ^{x'} W_\phi ^\alpha  (x,x') ] \\
  & = - Tr[U(x,x')\tilde{\gamma} ^\mu (x) \tilde{\gamma} ^\nu (x') U(x',x) \nabla _\mu ^{x} W_\phi ^\alpha  (x,x')  \nabla _\nu ^{x'} W_\phi ^\alpha  (x,x') ] \\
& =  - Tr[ U(x,x') b_\delta ^{.\mu} (x) {\gamma} ^\delta  b_\beta ^{.\nu} (x') \gamma ^\beta U(x',x) \nabla _\mu ^{x} W_\phi ^\alpha  (x,x')  \nabla _\nu ^{x'} W_\phi ^\alpha  (x,x') ] \numberthis 
\label{templabel}
\end{align*}
Here, we have used the relations \( A(x) \bar{A}(x') = U(x',x) \) and \( U(x',x) U(x,x') = \mathds{1} \) in the second last (sixth) line. Comparing the second last equation in Eq.(\ref{templabel}) with the last equation in Eq.(\ref{templabel2}), one can check that the ordering of the $U(x,x')$s and the $\tilde{\gamma} ^\mu (x)$ are different which leads to the 2-bein dependence in the final expression. Using $Tr[XY] = Tr[YX]$ and then again \( U(x',x) U(x,x') = \mathds{1} \) one can get rid of the spin parallel propagator as follows.
\begin{align*}
W_{\psi,b} ^{'\alpha} (x,x') & = - Tr[  b_\beta ^{.\nu} (x') \gamma ^\beta U(x',x) U(x,x') b_\delta ^{.\mu} (x) {\gamma} ^\delta  \nabla _\mu ^{x} W_\phi ^\alpha  (x,x')  \nabla _\nu ^{x'} W_\phi ^\alpha  (x,x') ] \\
& = - Tr[  b_\beta ^{.\nu} (x') \gamma ^\beta b_\delta ^{.\mu} (x) {\gamma} ^\delta  \nabla _\mu ^{x} W_\phi ^\alpha  (x,x')  \nabla _\nu ^{x'} W_\phi ^\alpha  (x,x') ] \\
& = - Tr[b_\beta ^{.\nu} (x')  b_\delta ^{.\mu} (x) \eta ^ {\delta \beta} \mathds{1} \partial _\mu ^x W_\phi ^\alpha  (x,x')  \partial _\nu ^{x'} W_\phi ^\alpha  (x,x')] \\
& = -Tr[C' \mathds{1}] \\
 & = - N b_\beta ^{.\nu} (x')  b_\delta ^{.\mu} (x) \eta ^ {\delta \beta} \partial _\mu ^x W_\phi ^\alpha  (x,x')  \partial _\nu ^{x'} W_\phi ^\alpha  (x,x') . \numberthis \label{ageneraltwoold}
\end{align*}
In the third line, we have used the anticommutator relation of the Dirac matrices, \( \{ \gamma^\delta, \gamma^\beta \} = 2 \mathds{1} \eta^{\delta\beta} \). Since all indices are contracted, we can again express everything other than \( \mathds{1} \) as \( C' \) and again use the identity \( \text{Tr}[C' \mathds{1}] = N C' \), where \( N \) is the dimensionality of spacetime, to obtain the final expression, Eq.~\eqref{ageneraltwoold}. One can notice that the Eq.~\eqref{ageneraltwoold} is now dependent on the 2-beins at $x$ and $x'$, because of the presence of $A$ and $\bar{A}$ in the transformed field $\psi'$ which cancels with the spin parallel propagator term.

\section{Response rate of a static detector in the HHI state } \label{Appendix B}
The pullback of the two-point function shown in Eq. (\ref{eq:13}) along the trajectory of a static detector is given by,
\begin{align*}
    W_{\psi,s} ^{'\text{HHI}} (x(\tau),x(\tau ^{'})) & =  -\frac{1}{4 \pi ^2 \sqrt{1-R_s/r}\sqrt{1-R_s/r'}}  \bigg[ e^{(v'-u)/2R_s} +  e^{(v-u')/2R_s} \bigg] \frac{1}{(\Delta U - i \epsilon)(\Delta V - i \epsilon)}\\
    & =  - \frac{1}{32 \pi ^2 R_S ^2 (1- \frac{R_S}{r}) } \frac{\cosh{(t_S'-t_S)/2 R_s}}{ \sinh ^2 {\bigg( \frac{t_S'-t_S-i\epsilon}{4 R_S}} \bigg)} .
\end{align*}
\\
Here, we have used \( U = -2R_s e^{-u/2R_s} \) and \( V = 2R_s e^{v/2R_s} \), where \( u = t_S - r_* \) and \( v = t_S + r_* \) with \( r = r' \) in the second line. For a static detector, \( d\tau = \sqrt{1 - R_s/r} \, dt_S = \kappa \, dt_S \) (say). Therefore, the above expression becomes

\begin{equation} \label{eq:62}
    W_{\psi,s} ^{'\text{HHI}} (x(\tau),x(\tau ^{'})) = - \frac{1}{32 \pi ^2 R_S ^2 \kappa ^2 } \frac{\cosh{\Delta \tau / 2 \kappa R_s}}{ \sinh ^2 {\bigg( \frac{ \Delta \tau -i\epsilon}{4 R_S \kappa}} \bigg)}
\end{equation}
\\
Using the periodicity in the imaginary time of the above expression one can see the KMS thermality of the HHI state. Substituting for $ W_{\psi,s} ^{'\text{HHI}}$ from Eq.\eqref{eq:62} in Eq.\eqref{eq:31}, we get the following transition rate

\begin{align*}
    \dot{\mathcal{F}} & = - \frac{1}{32 \pi ^2 R_S ^2 \kappa ^2 }  \int _{-\infty} ^{\infty} d\Delta \tau e^{-i \Omega \Delta \tau} \frac{\cosh{\Delta \tau /2 \kappa R_s}}{ \sinh ^2 {\bigg( \frac{ \Delta \tau -i\epsilon}{4 R_S \kappa}} \bigg)} \\
    & = - \frac{1}{32 \pi ^4 R_S ^2 \kappa ^2 } \sum _{k = -\infty} ^ {k= \infty} \int _{-\infty} ^{\infty} d\Delta \tau  \frac{e^{-i \Omega \Delta \tau} \cosh{\Delta \tau /2 \kappa R_s} }{\bigg(   \frac{ \Delta \tau - i \epsilon}{4 \pi R_S \kappa i} -k  \bigg) ^2} \\
    & = \frac{1}{2 \pi ^2} \sum _{k = -\infty} ^ {k= \infty} \int _{-\infty} ^{\infty} d\Delta \tau  \frac{e^{-i \Omega \Delta \tau} \cosh{\Delta \tau /2 \kappa R_s}}{\bigg(   \Delta \tau - i \epsilon - i 4 \pi k R_S \kappa  \bigg) ^2} ,
\end{align*}
\\
where we have used $\cosec^2{\pi x} = \pi ^{-2} \sum _{k = -\infty} ^ {k= \infty} (x-k)^{-2} $ in the second line. By choosing contour in the lower half plane to do this integration, we get 

\begin{eqnarray} \label{eq:63}
    \dot{\mathcal{F}}  =  \frac{\Omega}{\pi} \sum _{k = 1 } ^ {k= \infty} e^{-4 \pi k \kappa R_S \Omega}
    = \frac{\Omega}{\pi (e^ {4 \pi \kappa R_S \Omega } - 1)} .
\end{eqnarray}

\section{Response rate of a static detector in the Unruh state } \label{Appendix C }
The pullback of two-point function in Eq.\eqref{eq:17} along the trajectory of a static detector is given by,

\begin{align*} \numberthis
    W_{\psi,s} ^{'\text{Unruh}} (x,x') & = -\frac{1}{4 \pi ^2 \sqrt{1-R_s/r}\sqrt{1-R_s/r'}}  \bigg[ e^{-u/2R_s} +  e^{-u'/2R_s} \bigg] \frac{1}{(\Delta U - i \epsilon)(\Delta v - i \epsilon)} \\
    & = -\frac{1}{8 \pi ^2 R_s (1- \frac{R_s}{r}) } \frac{\cosh{\bigg( \frac{t_S'-t_S}{4 R_s} \bigg)}}{ \sinh {\bigg( \frac{t_S'-t_S-i\epsilon}{4 R_s}} \bigg) ( t_S'-t_S-i\epsilon )} .
\end{align*}
\\
Substituting the above expression in Eq.\eqref{eq:31}, we get the following transition rate:

 \begin{align*} \numberthis \label{equn79}
    \dot{\mathcal{F}} & = - \frac{1}{ 8 \pi ^2 R_s \kappa }  \int _{-\infty} ^{\infty} d\Delta \tau e^{-i \Omega \Delta \tau} \frac{\cosh{\bigg( \frac{\Delta \tau}{4 R_s \kappa } \bigg)}}{ \sinh {\bigg( \frac{\Delta \tau -i\epsilon}{4 R_s \kappa}} \bigg) ( \Delta \tau -i\epsilon )} \\
    & = \frac{1}{ 4 \pi ^2 R_s \kappa } \sum _{k = 1} ^ {\infty} \frac{e^{-4 \pi \kappa R_s \Omega n}}{n} .
\end{align*}
Here, we have used the contour in the lower half plane and summed all residues to write the second line.
Summing the infinite series one obtains

\begin{eqnarray} \label{eq:64}
    \dot{\mathcal{F}} = \frac{1}{4 \pi ^2 R_s \kappa} \log { \frac{e^ {4 \pi \kappa R_s \Omega }}{(e^ {4 \pi \kappa R_s \Omega } - 1)}} .
\end{eqnarray}

Using expansion of $\log({x+1})$ about x=1 we get the following series for the transition rate,

\begin{equation} \label{eq:65}
    \dot{\mathcal{F}} (\Omega) =  \frac{1}{4 \pi ^2 R_s \kappa } \bigg[ 
    \frac{1}{e^ {4 \pi \kappa R_s \Omega } - 1} -\frac{1}{2}  \bigg(  \frac{1}{e^ {4 \pi \kappa R_s \Omega } - 1}  \bigg) ^2  + \frac{1}{3}  \bigg(  \frac{1}{e^ {4 \pi \kappa R_s \Omega } - 1}  \bigg) ^3 + ......  \bigg] .
\end{equation}
Repeating the above calculation with the 2-bein \eqref{conftetru} one gets similar expression in Eq.\eqref{equn79}, without $\cosh$ factor in the nominator. Therefore, we get the transition rate
\begin{align*} \numberthis
    \dot{\mathcal{F}} = \frac{1}{ 4 \pi ^2 R_s \kappa } \sum _{k = 1} ^ {\infty} \frac{(-1)^{n+1} e^{-4 \pi \kappa R_s \Omega n}}{n} .
\end{align*}
Performing the series sum in the above expression, we get 
\begin{eqnarray} \label{eq:82}
    \dot{\mathcal{F}} (\Omega) = \frac{1}{4 \pi ^2 R_s \kappa} \log { \frac{e^ {4 \pi \kappa R_s \Omega }}{(e^ {4 \pi \kappa R_s \Omega } + 1)}}.
\end{eqnarray}

\section{Computing transition probability} \label{Appendix D}
In this appendix, we apply the saddle point approximation to compute the diagonal terms of $\rho_{AB}$, known as the transition probability \cite{saddle, ARFKEN2013551}. The transition probability is given by the following type of integration
\begin{align*}
     L  = & \lambda ^2  \int _{-\infty} ^{+\infty} d\tau \int _{-\infty} ^{+\infty} d\tau' \chi (\tau) \chi(\tau') e^{- i \Omega (\tau - \tau')} W_\psi ^{\alpha} (x(\tau),x(\tau ')) \\
      = &   \lambda ^2 \int _{-\infty} ^{+\infty} d\tau \int _{-\infty} ^{+\infty} d\tau' e^{- \frac{(\tau - \tau _0)^2}{2 \sigma ^2}} e^{- \frac{(\tau ' - \tau _0)^2}{2 \sigma ^2}} e^{- i \Omega (\tau - \tau')} W_\psi ^{\alpha} (x(\tau),x(\tau ')) \\
       = &   \lambda ^2 e^{-\sigma ^2 \Omega ^2} \int _{-\infty} ^{+\infty} d\tau \int _{-\infty} ^{+\infty} d\tau' e^{- \frac{(\tau - \tau _0 +  i \Omega \sigma ^2 /2)^2}{2 \sigma ^2}} e^{- \frac{(\tau ' - \tau _0 - i \Omega \sigma ^2 /2)^2}{2 \sigma ^2}} W_\psi ^{\alpha} (x(\tau),x(\tau ')).
\end{align*}
Now we use the fact that in the case of static detector $W_\psi ^{\alpha} (x(\tau),x(\tau '))$ is a function of $\tau - \tau '$ viz, it depends only on the interval of time. Changing variables from $\tau, \tau'$ to $\tilde{x}$  = $\tau + \tau'$
and $\tilde{y}$ = $\tau - \tau'$   one gets
\begin{eqnarray*}
    L = &   \lambda ^2 e^{-\sigma ^2 \Omega ^2} \int _{-\infty} ^{+\infty} d\tilde{x} \int _{-\infty} ^{+\infty} d\tilde{y} e^{- \frac{(\tilde{y} +  i 2 \Omega \sigma ^2 )^2}{4 \sigma ^2}} e^{- \frac{(\tilde{x} - 2 \tau_0)^2 }{4 \sigma ^2}} W_\psi ^{\alpha} (x(\frac{\tilde{x}+\tilde{y}}{2}),x(\frac{\tilde{x}-\tilde{y}}{2})).
\end{eqnarray*}
\\
Shifting the contour in an imaginary direction for $\tilde{y}$  by $ 2 \Omega \sigma ^2 $  makes the exponential factors real. Now, the saddle point approximation on the resulting expression gives \cite{saddle, ARFKEN2013551} 
\begin{equation} \label{eq:66}
    L \approx 2 \pi \sigma ^2 \lambda ^2 e^{- \sigma ^2 \Omega ^2} W_\psi ^{\alpha} (x(\tau_0 - i \Omega \sigma^2),x(  \tau_0 + i \Omega \sigma^2))  + \text{ residue terms}.
\end{equation}
\\
The residue term that comes from shifting the contour vanishes if one restricts to $\Omega \sigma ^2 / 2\kappa R_s < \pi$, as no pole is being crossed.

\section{Monotonicity of $ h(x)$ and $g(x)$} \label{Appendix E}
Let us define a function $h(x)$ as
\begin{equation} \label{defineh}
    h(x) = \frac{x_2 - x}{e^{-x_2} - e^{-x}} \frac{e^{-x_1} - e^{-x}}{x_1 - x}.
\end{equation}
The sign of the denominator of the first derivative of the above function is positive since we get a square after differentiation. Therefore, the overall sign is determined by the numerator, which is given by
\begin{align*}
   \text{sign} (h'(x)) & =\text{sign} [ (x_2 - x_1)(e^{-x_1 - x_2} + e^{- 2 x}) + (x_1 x_2 - (x_1 + x_2)x + x^2 + x_1- x_2)e^{-x - x_2}  \\ & -  (x_1 x_2 - (x_1 + x_2)x + x^2 + x_2- x_1)e^{-x - x_1}  ] \\ 
   & =  \text{sign} [ (x_2 - x_1)(e^{x -x_1} + e^{-x + x_2}) + (x_1 x_2 - (x_1 + x_2)x + x^2 + x_1- x_2)\\ & -  (x_1 x_2 - (x_1 + x_2)x + x^2 + x_2- x_1)e^{x_2 - x_1}  ] . \numberthis \label{withoutas1}
\end{align*}
Assuming the points $x_1$ and $x_2$ are close together, we can ignore quadratic and higher-order terms in the expansion of $e^{x_2 - x_1}$. Keeping terms linear in $x_2 - x_1$, we obtain:
\begin{align*}
    \text{sign}(h'(x)) = \text{sign}\left[ (x_2 - x_1)\left(e^{x - x_1} + e^{-x + x_2} - x^2 + (x_2 + x_1)x - x_1 x_2 -x_2 + x_1 - 2\right) \right].
\end{align*}
If $x_1 > x_2 > x$ and if $x_1$ and $x_2$ are sufficiently larger than $x$, then $e^{-x + x_2}$ will be dominant and we will have
\begin{equation} \label{mono1}
    \text{sign}\left( e^{x - x_1} + e^{-x + x_2} - x^2 + (x_2 + x_1)x - x_1 x_2  - x_2 + x_1 - 2 \right) > 0.
\end{equation}
This implies $h'(x) < 0$. In other words, under these assumptions, $h(x)$ is a monotonically decreasing function of $x$. Furthermore, one can see that if \( x_1 < x_2 < x \), and \( x_3 \) and \( x_4 \) are sufficiently larger than \( x_2 \), then the \( e^{x - x_2} \) term will be dominant. The terms in the second bracket, Eq. \eqref {mono1}, as well as the first term, will be positive, which implies that \( h'(x) > 0 \). In this case $h(x)$ will be a monotonically increasing function of $x$.
\begin{comment}
Using the inequality \( e^{-x} - e^{-x_4} > (x_4 - x) e^{-x} \), or equivalently, \( e^{x - x_4} < 1 + x - x_4 \), we can conclude that \( h'(x) > 0 \) for all \( x \). Therefore, \( h(x) \) is a monotonically increasing function of \( x \). As  a result, if $x_4 > x_3 > x2 > x_1$ then
\begin{equation}
    \frac{x_1 - x_4}{e^x - e^{x_4}} <  \frac{x_2 - x_4}{e^x - e^{x_4}}
\end{equation}
\end{comment}

Further, let us define another function g(x) as
\begin{equation} \label{defineg}
    g(x) = \frac{x_2 - x}{e^{x_2} - e^{x}} \frac{e^{x_1} - e^{x}}{x_1 - x}.
\end{equation}
The sign of the denominator of the first derivative of the above function is also positive since we get a square after differentiation. Therefore, the overall sign is determined by the nominator, which is given by
\begin{align*}
   \text{sign} (g'(x)) & =\text{sign} [ (x_2 - x_1)(e^{x_1 + x_2} + e^{ 2 x}) + (- x_1 x_2 + (x_1 + x_2)x - x^2 + x_1- x_2)e^{x + x_2}  \\ & +  (x_1 x_2 - (x_1 + x_2)x + x^2 - x_2 + x_1)e^{x + x_1}  ] \\ 
   & =  \text{sign} [ (x_2 - x_1)(e^{x_1 -x} + e^{x - x_2}) + (- x_1 x_2 + (x_1 + x_2)x - x^2 + x_1- x_2)\\ & +  (x_1 x_2 - (x_1 + x_2)x + x^2 + x_1- x_2)e^{x_1 - x_2}  ] \numberthis \label{withoutas1}
\end{align*}
Once again, assuming the points $x_1$ and $x_2$ to be close together, we can ignore quadratic and higher-order terms in the expansion of $e^{x_2 - x_1}$. Keeping terms linear in $x_2 - x_1$, we obtain:
\begin{align*}
    \text{sign}(g'(x)) = \text{sign}\left[ (x_2 - x_1)\left(e^{x - x_2} + e^{-x + x_1} - x^2 + (x_2 + x_1)x - x_1 x_2 -x_1 +x_2 - 2\right) \right].
\end{align*}
If $x_1 < x_2 < x$ and if x is sufficiently larger than $x_1$ and $x_2$, then $e^{x - x_2}$ will be dominant and we will have
\begin{equation} \label{mono2}
    \text{sign}\left(e^{x - x_2} + e^{-x + x_1} - x^2 + (x_2 + x_1)x - x_1 x_2 -x_1 +x_2 - 2\right) > 0.
\end{equation}
This implies $g'(x) > 0$. In other words, under these assumptions, $g(x)$ is a monotonically increasing function of $x$.  Furthermore, one can see that if \( x_1 > x_2 > x \), and \( x_1 \) and \( x_2 \) are sufficiently larger than \( x \), then the \( e^{-x + x_1} \) term will be dominant. The terms in the second bracket, Eq.\eqref{mono2}, will be positive while the term in the second bracket will be negative, which implies that \( g'(x) < 0 \). In this case, $g(x)$ will be a monotonically decreasing function of $x$.

\end{appendices}

\bibliography{fermion}

\end{document}